\documentclass[preprint,12pt,authoryear]{elsarticle}

\usepackage{graphicx}
\usepackage{amsmath}
\usepackage{amssymb}
\usepackage{amsthm}

\usepackage{caption}
\usepackage{subcaption}
\usepackage{newtxtext}
\usepackage{newtxmath}
\usepackage{hyperref}
\hypersetup{
    colorlinks = true,
    urlcolor   = blue,
    citecolor  = black,
}
\usepackage[svgnames]{xcolor}

\usepackage{ulem}

\usepackage{multicol}
\usepackage{multirow}
\usepackage{adjustbox}

\usepackage{subcaption}
\expandafter\def\csname ver@subfig.sty\endcsname{}
\usepackage{svg}
\usepackage{hyperref}

\usepackage{tikz}
\newcommand{\solidline}{\raisebox{0.6mm}{\tikz{\draw[-,black,solid,line width = 0.4mm](0,0) -- (3.5mm,0);}}}
\newcommand{\dashedline}{\raisebox{0.6mm}{\tikz{\draw[-,black,dashed,line width = 0.4mm](0,0) -- (3.5mm,0);}}}
\newcommand{\dashdotted}{\raisebox{0.6mm}{\tikz{\draw[-,black,dashdotted,line width = 0.4mm](0,0) -- (3.5mm,0);}}}
\newcommand{\densdashdot}{\raisebox{0.6mm}{\tikz{\draw[-,black,densely dashdotted,line width = 0.4mm](0,0) -- (3.5mm,0);}}}

\usepackage[ruled, linesnumbered]{algorithm2e}
\SetKwInput{KwInput}{Input}                
\SetKwInput{KwOutput}{Output}              
\usepackage{setspace} 

\begin{document}

\begin{frontmatter}



\title{Embedded shear layers in turbulent boundary layers \\ of a NACA0012 airfoil at high angles of attack}



\author[inst1]{Leandro J. O. Silva}
\author[inst1]{William R. Wolf}

\affiliation[inst1]{organization={Faculdade de Engenharia Mecânica, Universidade Estadual de Campinas},
            city={Campinas},
            postcode={13086-860},
            country={Brazil}}

\begin{abstract}
An investigation of turbulent boundary layers (TBLs) is presented for a NACA0012 airfoil at angles of attack 9 and 12 deg. Wall-resolved large eddy simulations (LES) are conducted for a freestream Mach number $M = 0.2$ and chord-based Reynolds number $Re = 4\times 10^5$, where the boundary layers are tripped near the airfoil leading edge on the suction side. For the angles of attack analyzed, mild, moderate and strong adverse pressure gradients (APGs) develop over the airfoil. Despite the strong APGs, 
the mean flow remains attached along the entire airfoil suction side. Similarly to other APG-TBLs investigated in the literature, a secondary peak appears in the Reynolds stress and turbulence production profiles. This secondary peak arises in the outer layer and, for strong APGs, it may overcome the first peak typically observed in the inner layer. The analysis of the turbulence production shows that other components of the production tensor become important in the outer layer besides the shear term. 
For moderate and strong APGs, the mean velocity profiles depict three inflexion points, the third being unstable under inviscid stability criteria. In this context, an embedded shear layer develops along the outer region of the TBL leading to the formation of two-dimensional rollers typical of a Kelvin-Helmholtz instability which are captured by a spectral proper orthogonal decomposition (SPOD) analysis. The most energetic SPOD spatial modes of the tangential velocity show that streaks form along the airfoil suction side and, as the APG becomes stronger, they grow along the spanwise and wall-normal directions, having a spatial support along the entire boundary layer. 
\end{abstract}



\begin{keyword}
Turbulent boundary layer \sep adverse pressure gradient \sep embedded shear layer \sep high angle of attack \sep large eddy simulation 
\end{keyword}

\end{frontmatter}

\section{Introduction}

Airfoil profiles are employed in wings, wind turbines, rotorcraft, and propellers to generate lift or thrust. In these applications, the operational envelopes may span a broad range of angles of attack, and the boundary layers, which are typically turbulent, develop under different pressure gradients. Early experimental studies of turbulent boundary layers (TBLs) subjected to adverse pressure gradients (APGs) were conducted by \citet{clauser1954,clauser1956}. In these references, the author investigated the relevant flow parameters that provided a scaling for suitable equilibrium boundary layer profiles. The pressure gradient parameter $\beta = (\delta^*/\tau_w) dp/dx$ was defined and it would be later known as the Clauser parameter. Here, $\delta^*$ represents the boundary layer displacement thickness, $\tau_w$ is the wall shear stress, and the streamwise pressure gradient is given by $dp/dx$.

Another study that contributed to paving the way for understanding the influence of pressure gradients on TBLs was presented by \citet{bradshaw1967}, who conducted measurements of boundary layers with zero pressure gradient (ZPG), as well as with moderate and strong APGs. It was demonstrated that for stronger APGs, large eddy motions in the outer layer increased in strength, contributing significantly to the shear stress and, subsequently, to the production of turbulent kinetic energy (TKE). Such turbulent structures were considered similar to those found in free mixing layers.

Experiments with increasing APG distributions along a flat plate were performed by \citet{samuel1974} who observed the development of more pronounced outer regions of the  mean velocity profiles with the flow development when the standard inner-wall scaling was applied. A similar observation was made by \citet{nagano1993}, who conducted experiments on a flat plate with moderate and strong APG-TBLs. They noticed that the APGs affected the outer region of the boundary layer both in terms of the Reynolds stresses and the logarithmic layer of the mean velocity profiles. 
In the same year, \citet{spalart1993} performed  experiments and direct numerical simulations (DNS) of TBLs developing under different values of the Clauser parameter. Through a comparison of results, they verified that the logarithmic region of the inner-scaled mean velocity profile exhibited a vertical downward shift in cases with APGs. \citet{skaare1994} conducted experiments with strong APGs and showed that the production term of the TKE budget presents a second peak in the outer region of the boundary layer, related to the high shear stresses in this region. Furthermore, they also observed that the dissipation term was higher than typically found for a ZPG-TBL, being significant both near and away from the wall. Complementarily, \citet{krogstad1995} performed a quadrant decomposition analysis of the Reynolds shear stress for an APG-TBL. They showed that, near the wall, the flow was dominated by strong events in the fourth quadrant, i.e., by turbulent motions directed towards the wall. 

Advances in numerical simulations and experimental techniques allowed evaluating not only the turbulence statistics, but also the characterization of coherent flow structures in TBLs. \citet{skote2002} performed DNS of TBLs subjected to strong APGs. While in one of the cases simulated a separation bubble was formed, in the other the boundary layer remained attached. They showed that the near-wall streaks were weakened under a strong APG. This observation was also made by \citet{lee2008,lee2009}, who conducted DNS of TBLs subjected to different APG conditions and compared the results with a ZPG flow. The authors commented that the outer layer peak production in the TKE budget analysis was related to hairpin-like vortices enhanced by the APG. They also highlighted the fact that the standard logarithmic law of the wall is not valid for APG flows. In addition, experiments were conducted by \citet{monty2011} and compared to a database composed of other experimental and numerical results in order to perform a parametric study of APG effects in TBLs. In this case, it was shown that the large-scale structures in the TBL were energized due to the APG, resulting in a higher turbulence intensity. The authors also observed the same effect when the Reynolds number was increased while maintaining the same pressure gradient. However, the energy amplifications due to variations in the APGs were higher than those from the increase in the Reynolds number.

To better understand the modification of the large-scale motions in TBLs subjected to pressure gradients, \citet{harun2013} performed experiments of boundary layers under favorable and adverse pressure gradients. The authors observed that the outer region of the cases analyzed were significantly different in terms of the turbulence intensity and production. Their spectral analysis showed that the large-scale motions are amplified in APG-TBLs, especially in the outer region, whereas such motions are attenuated when subjected to favorable pressure gradients (FPGs). \citet{schiavo2015, schiavo2017} studied the development of turbulent boundary layers in a convergent-divergent channel, and the effects of FPGs and APGs were investigated through budgets of TKE and the individual components of the Reynolds stresses. Proper orthogonal decomposition (POD) was also applied to reconstruct the flows using a percentage of the total kinetic energy to understand the role of the most energetic structures in the TKE budgets. The results showed that these structures account for most of the terms appearing in the budget, with exception of the turbulent transport. Furthermore, results from spectral analysis indicated that, for the APG cases, the TKE was transported both towards the wall and the channel center. 

A complementary perspective about the APG effects in TBLs was given by \citet{schatzman2017}. In their work, experiments were conducted for an unsteady APG-TBL where the pressure gradient was time dependent, leading to a cycle of separation and reattachment. The results demonstrated that when the boundary layer was exposed to an APG, an inflectional point originated in the mean velocity profile indicating an inviscid instability associated with the existence of an embedded shear layer. The authors also showed that strong sweep events were the dominant contributors to the Reynolds stresses in the near-wall region, whereas away from the wall, ejections became  dominant. This was shown to be different from ZPG-TBLs, where ejection events are the major contributors to the Reynolds stresses along the entire boundary layer. They also observed that the location of the peak ejection (sweep) events occur in the higher (lower) velocity region of the embedded shear layer. 
More recently, \citet{balantrapu2023} performed an experiment of a body of revolution in a high Reynolds number flow with a strong APG. Their results revealed the existence of coherent structures with coupled negative and positive peaks, indicating convective rollers.

Extensive studies have been performed to investigate APG-TBLs due to airfoil camber effects \citep{hosseini2016,vinuesa2017}. 
These authors showed that the buffer layer
is affected for moderate values of the APG, suggesting a different momentum transport mechanism through the boundary layer related to the more intense large-scale motions. Differences in the energy distribution were also discussed in the previous references, where the TKE production presents a secondary peak in the outer layer for strong APGs, and the dissipation of TKE is increased along the entire boundary layer. History effects were also studied in order to assess the influence of the streamwise pressure gradients by analyzing several databases of APG-TBLs developing over
flat plates and wings  \citep{bobke2017, vinuesa2017rev}. In these studies, it was shown that the mean velocity and Reynolds stress profiles are dependent of the flow development. Furthermore, \citet{vinuesa2018} performed large eddy simulations (LES) of a NACA4412 with chord-based Reynolds numbers ranging from $Re = 1\times10^5$ to $1\times10^6$. The simulations allowed understanding the mechanisms responsible for the development of the outer region of TBLs. Their results demonstrated that there are two complementing mechanisms, one related to the increase in Reynolds number and another associated with the APG. They also showed that low Reynolds number boundary layers are more affected by the APGs due to the wall-normal convection. This would in turn thicken the boundary layers, enabling the formation of larger outer regions and more energetic large-scale motions. 
\citet{tanarro2020} performed wall-resolved LES of NACA0012 and NACA4412 airfoils at 0 and 5 deg. angles of attack, and compared results with a ZPG-TBL. It was observed that the turbulence statistics are considerably affected by the APG in the outer layer. Their solutions were further analyzed by power-spectral density maps which demonstrated that both the large and small scales were energized in the outer region, a different behavior compared to ZPG-TBLs.
This suggests that the APG induces a transport of small scales from the near-wall to the outer region, confirming that the energization mechanisms due to the APGs are different from those of high Reynolds numbers.

\citet{maciel2018} analyzed experimental and numerical databases of TBLs with different APG conditions developing on divergent channels, a wing and a ZPG-TBL. They established a set of nondimensional parameters to characterize the outer region of a TBL subjected to an APG. It was shown that the best scaling was obtained in terms of the boundary layer integral quantities as well as the velocity at the edge of the boundary layer.
More recently, \citet{wei2023} proposed a new scaling of the mean momentum equation for the outer region of an APG-TBL. Variables typically used in shear layer scalings were employed and results were compared using a database of experiments and simulations of channel flows and flat plates. A good scaling agreement was observed using the mean velocity defect and Reynolds stresses along the outer layer. In summary, the previous results support the idea that an embedded shear layer may be a feature present in APG-TBLs, as also suggested by \citet{bradshaw1967} and \citet{schatzman2017}.

In the present work, we study the effects of APGs on TBLs developing over a NACA0012 airfoil at high angles of attack, but without mean flow separation. 
Wall-resolved LES are performed for a NACA0012 profile at angles of attack $9$ and $12$ deg.  The Reynolds and Mach numbers are set as $Re = 4 \times 10^5$ and $M=0.2$, respectively, and tripping is enforced near the leading edge to guarantee the development of fully turbulent boundary layers along the airfoil suction side. The following sections are organized as follows: in section 2, the numerical methodology is presented also including the flow conditions investigated as well as the grid configurations. Section 3.1 presents results in terms of integral quantities. Then, turbulence statistics are computed including an analysis of the mean flow profiles (section 3.2), Reynolds stresses (section 3.3), and TKE budgets (section 3.4). The turbulence production term is investigated in further details in section 3.5, including its individual components and spatial distribution. An assessment in terms of flow anisotropy is presented in section 3.6 by analyzing trajectories in the Lumley triangle in conjunction with the individual components of the normalized anisotropy tensor. Finally, an inspection of coherent structures is presented in section 3.7 using a spectral proper orthogonal decomposition (SPOD), followed by the conclusions in section 4.

\section{Numerical methodology}

\subsection{Wall-resolved large eddy simulations}

Wall-resolved large eddy simulations are performed to solve the non-dimensional compressible Navier Stokes equations in general curvilinear coordinates. The spatial discretization of the governing equations is performed using a sixth-order accurate compact scheme implemented on a staggered grid \citep{nagarajan2003}. An overset grid procedure is employed, where a body-fitted O-grid conforms to the airfoil, while a Cartesian H-grid is used to enclose the entire computational domain. In the O-grid, the time integration is carried out using the implicit second-order scheme of \citet{beam1978} in order to overcome the stiffness problem of the fine near-wall grid resolution, whereas a third-order Runge-Kutta scheme is applied on the Cartesian mesh. For the communication between the grid overlapping zones, a fourth-order Hermite interpolation scheme is used \citep{bhaskaran2010}. 

In the present simulations, no explicit subgrid scale model is employed. However, a sixth-order compact filter \citep{lele1992} is applied away from the walls to control high-wavenumber numerical instabilities arising from grid stretching and interpolation between the grid blocks. The transfer function associated with such filters has been shown to provide an approximation to subgrid scale models \citep{mathew2003}.
No-slip adiabatic wall boundary conditions are enforced along the airfoil surface. At the farfield, characteristic boundary conditions based on Riemann invariants are employed together with a sponge layer that prevents reflection of acoustic waves. Periodic boundary conditions are applied in the spanwise direction.
The present numerical procedure has been validated for various simulations of compressible airfoil flows at different configurations \citep{wolf2012,wolf2012DU,brener2019,ricciardi2022prf,miotto2022,lui2022prf}, and further details on the numerical schemes employed in the current simulations are presented by \citet{nagarajan2003,bhaskaran2010} and \citet{wolf2011}. 

\subsection{Flow conditions and grid configurations}

The effects of different APGs on TBLs are assessed by increasing the angle of attack of a NACA0012 airfoil. Two simulations are performed, one for an incidence of $9$ deg. and another for $12$ deg. angle of attack. 
In both cases, the Reynolds number is set as $Re = 4 \times 10^5$ based on the inflow velocity $U_\infty$ and chord length $L^*$, and the freestream Mach number is $M = 0.2$. The angles of attack are chosen so that the boundary layer remains attached on the airfoil suction side.
The O-grid is generated in order to obtain smooth metric terms, and the NACA0012 airfoil is truncated at $99\%$ of the chord. Its modified trailing edge has a curvature radius of $r = 0.0015 L^*$. The leading edge is positioned at $(x,y) = (0,0)$ and the airfoil is pivoted about this point. The airfoil span is chosen to resolve at least 5 times the length of the boundary layer displacement thickness at the trailing edge in order to minimize the effects of the spanwise periodic boundary conditions.

\begin{figure}
  \centerline{\includegraphics[width=1\textwidth]{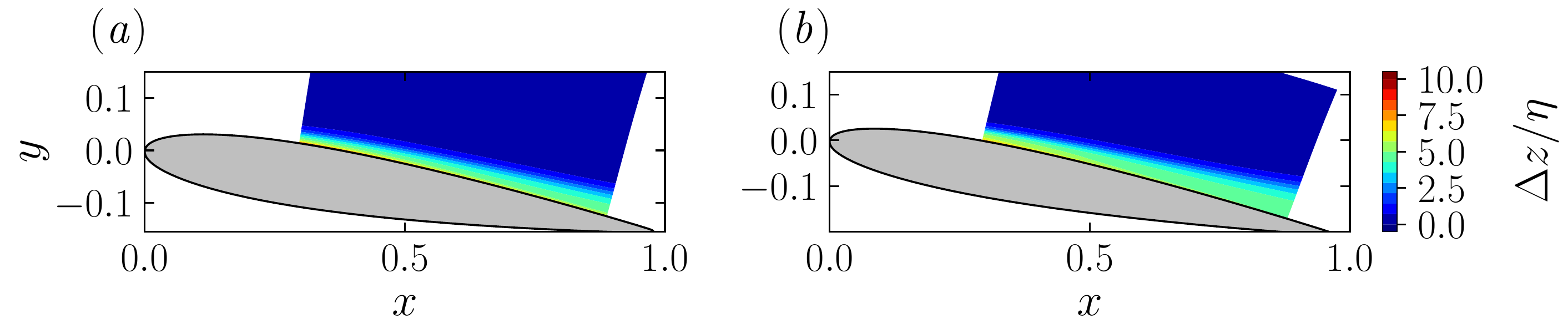}}
  \caption{Ratio of local spanwise grid size to the estimated Kolmogorov scale ($\Delta z / \eta$) for angle of attack of (\textit{a}) 9 and (\textit{b}) 12 deg.}
\label{fig:gridres}
\end{figure}

In order to verify the grid quality, two analyses are performed being one with respect to the near-wall resolution in terms of wall units, and another with respect to the ratio of the local spanwise grid size to the estimated Kolmogorov scale ($\Delta z / \eta$).   
Table \ref{tab:grid} presents the number of grid points ($N_x \times N_y \times N_z$) for each mesh used in the simulations as well as the respective spatial resolution in terms of wall units considering specific chord regions which serve as reference positions for the forthcoming analyses. As can be seen, the present grid resolutions show compliance with the values recommended by \citet{georgiadis2010} for a wall-resolved LES. The table also shows the spanwise domain employed for each angle of attack.
Figure \ref{fig:gridres} shows the values of $\Delta z / \eta$ over the suction side boundary layer for both cases studied. These values are computed based on the local turbulence dissipation and, as can be observed, $\Delta z / \eta < 10$ everywhere. These values confirm the quality of the present grids, and the overall resolution is slightly lower than that of the DNS from \citet{marquillie2011}, being comparable to the wall-resolved LES of \citet{schiavo2015}.
The Cartesian H-grid block is generated with dimensions $-3.4 \le x \le 5.9$ and $-4.2 \le y \le 5.7$ such that there is no flow confinement effects due to the size of the computational domain.


\begin{table}
  \begin{center}
\def~{\hphantom{0}}
  \begin{tabular}{ccccccccc}
AoA & \multicolumn{2}{c}{Grid type and size} & Span & x & $\Delta x^+$ & $\Delta y^+$ & $\Delta z^+$ & Span/$\delta^*$ \\
\multirow{3}{*}{9 deg.} & O-grid & $1200 \times 170 \times 144$ & \multirow{3}{*}{$0.12 L^*$} 
& 0.5 & 37.7 & 0.4 & 14.5 & 23.2            \\
    & H-grid & $960 \times 599 \times 72$   &                             & 0.7 & 26.9 & 0.4 & 12.6 & 15.1 \\
                & & & & 0.9 & 9.9 & 0.3 & 9.7 & 8.4 \\
\multirow{3}{*}{12 deg.} & O-grid & $1200 \times 170 \times 288$ & \multirow{3}{*}{$0.24L^*$}  
& 0.5 & 31.4 & 0.4 & 12.2 & 25.0            \\
    & H-grid & $960 \times 599 \times 144$   &                             & 0.7 & 19.7 & 0.3 & 9.3 & 14.4 \\
                & & & & 0.9 & 5.3 & 0.2 & 5.3 & 7.4           
\end{tabular}
  \caption{Grid configuration and near-wall resolution in wall units for different chord positions. The number of grid points is shown as $N_x \times N_y \times N_z$ for the streamwise, wall-normal, and spanwise directions, respectively.}
  \label{tab:grid}
  \end{center}
\end{table}

\begin{figure}
  \centering
    \begin{subfigure}{0.49\textwidth}
    \includegraphics[width=\textwidth]{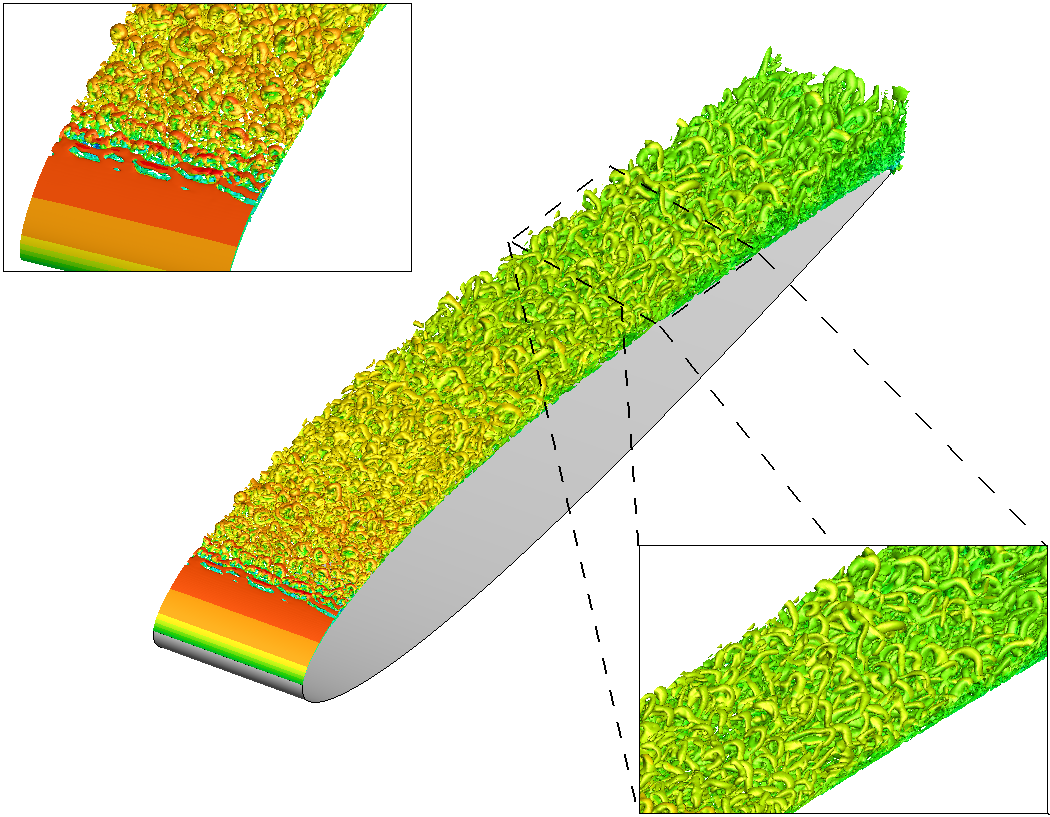}
    \caption{}
    \label{fig:snap_aoa9}
    \end{subfigure}
    \begin{subfigure}{0.49\textwidth}
    \includegraphics[width=\textwidth]{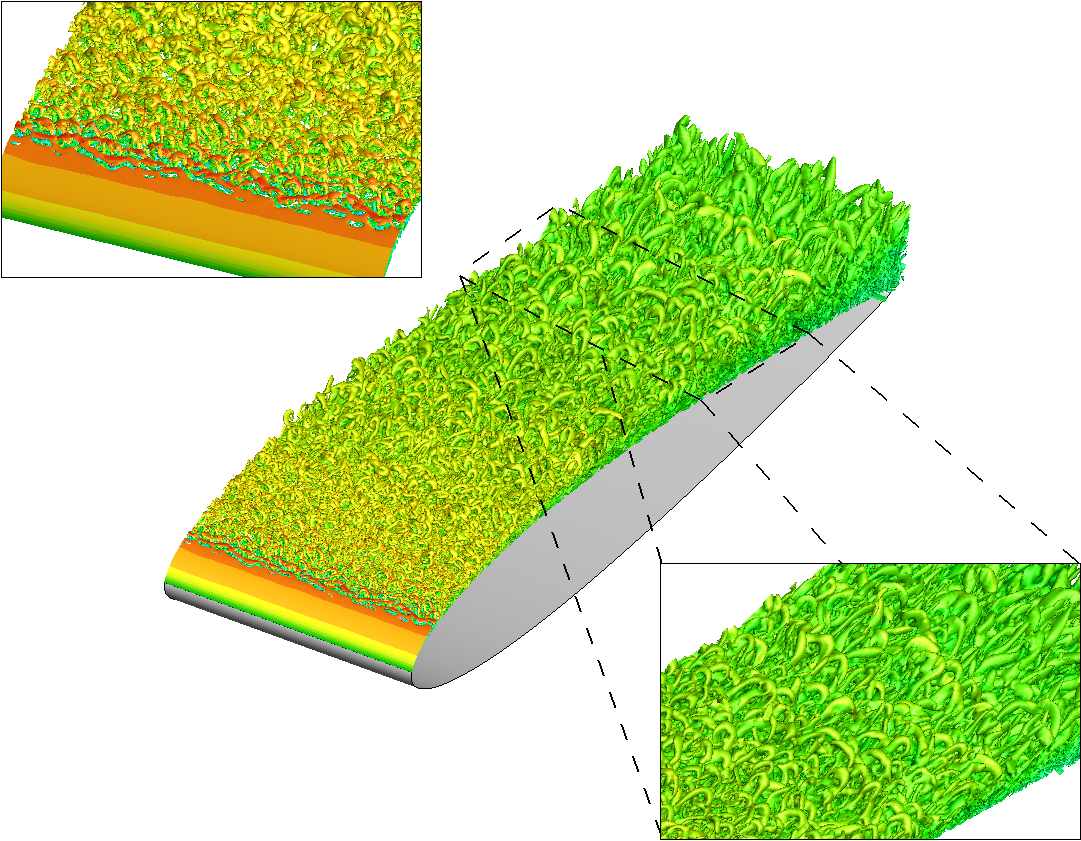}
    \caption{}
    \label{fig:snap_aoa12}
    \end{subfigure}
  \caption{Iso-surfaces of Q-criterion colored by $x$-momentum for (\textit{a}) $9$ and (\textit{b}) $12$ deg. Boundary layer tripping is applied on $0.04\leq x \leq 0.09$. 
  }
\label{fig:snapshots}
\end{figure}

A numerical tripping is enforced on both simulations from $0.04\leq x \leq 0.09$ on the suction side, which is the region where natural transition initiates. A random spanwise and streamwise tripping is applied in this region in order to avoid the presence of Tollmien-Schlichting-like waves that would otherwise appear. The tripping consists of blowing and suction which excite several wavenumbers with random phase variations. Zero-net-mass-flux is enforced and the maximum amplitude of the tripping is chosen so that a bypass transition occurs with a minimal disturbance to the flow. Results of the tripped boundary layer can be seen in figure \ref{fig:snapshots}, where fine turbulence scales are shown along the airfoil suction side for both simulations. In the figure, insets are also shown highlighting the hairpins formed along the TBLs. A clear difference is noticed in the length scales of these structures, where smaller (larger) scales are observed near the leading (trailing) edges.

\section{Results}

In order to investigate the angle of attack effects on TBLs including their impact on the APGs, post-processing of the LES results is performed along the airfoil suction side. Results are evaluated in terms of integral quantities and turbulence statistics, besides flow anisotropy and spectral proper orthogonal decomposition (SPOD).
The integral quantities provide information about the flow history effects. The consequences of mild, moderate and strong APGs are verified on different turbulence statistics, including the mean velocity, Reynolds stress and TKE budget profiles. A study in terms of the flow anisotropy allows a characterization of the turbulence states at different chord locations, i.e., under different APGs, in the wall-normal direction. Finally, the SPOD analysis provides an assessment of the most energetic coherent  structures along the boundary layer, including a visualization of their spatial support. 

\subsection{Integral quantities}

\begin{figure}
  \centerline{\includegraphics[width=0.90\textwidth]{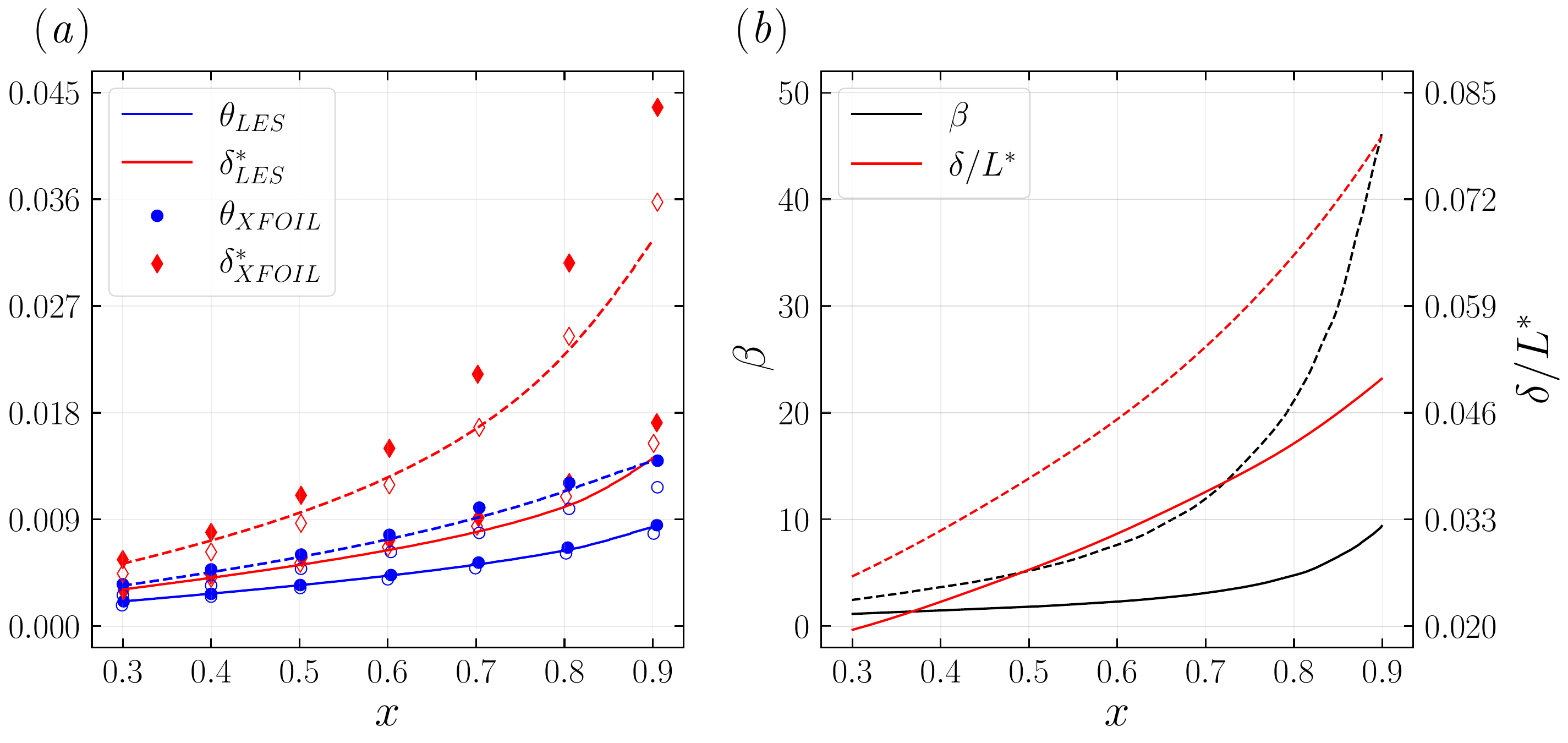}}
  \caption{Distribution of (\textit{a}) displacement and momentum thickness, and (\textit{b}) boundary layer thickness and Clauser's pressure gradient parameter along the suction side for $9$ deg. (solid lines) and $12$ deg. (dashed lines). The filled and open symbols represent results obtained from Xfoil for tripped boundary layers with $N-$factors $9$ and $5$, respectively.}
\label{fig:int_quant}
\end{figure}

In this section, boundary layer integral quantities are evaluated in terms of displacement thickness $\delta^*$, momentum thickness $\theta$, and the Clauser pressure gradient parameter $\beta$. These parameters require the calculation of the local boundary layer thickness $\delta$, which is computed by the approach proposed by \citet{vinuesa2016}. 
Figure \ref{fig:int_quant}(\textit{a}) presents the chordwise evolution of the nondimensional displacement and momentum thicknesses for both simulations. Results are compared with values obtained from Xfoil \citep{xfoil} for boundary layers with $N-$factors 9 (solid symbols) and 5 (open symbols) representing flows with small disturbances and bypass transition, respectively. It is important to remind that the LES calculations are tripped near the leading edge, in the region where TS-waves appear. In Xfoil, tripping is enforced in the same location. A good agreement is observed in terms of momentum thickness for both cases, and for the displacement thickness of the 9 deg. case. For 12 deg. angle of attack, the displacement thickness shows a good comparison with the forced transition case of $N=5$.
The Clauser pressure-gradient parameter is presented in figure \ref{fig:int_quant}(\textit{b}), where the difference between the APGs can be observed due to the variation in the angle of attack from 9 to 12 degs. For the higher incidence case, a steep increase in the pressure gradient occurs as the flow reaches the trailing edge region. In the same plot, the boundary layer thickness is also shown normalized by the chord.

\begin{figure}
  \centerline{\includegraphics[width=1\textwidth]{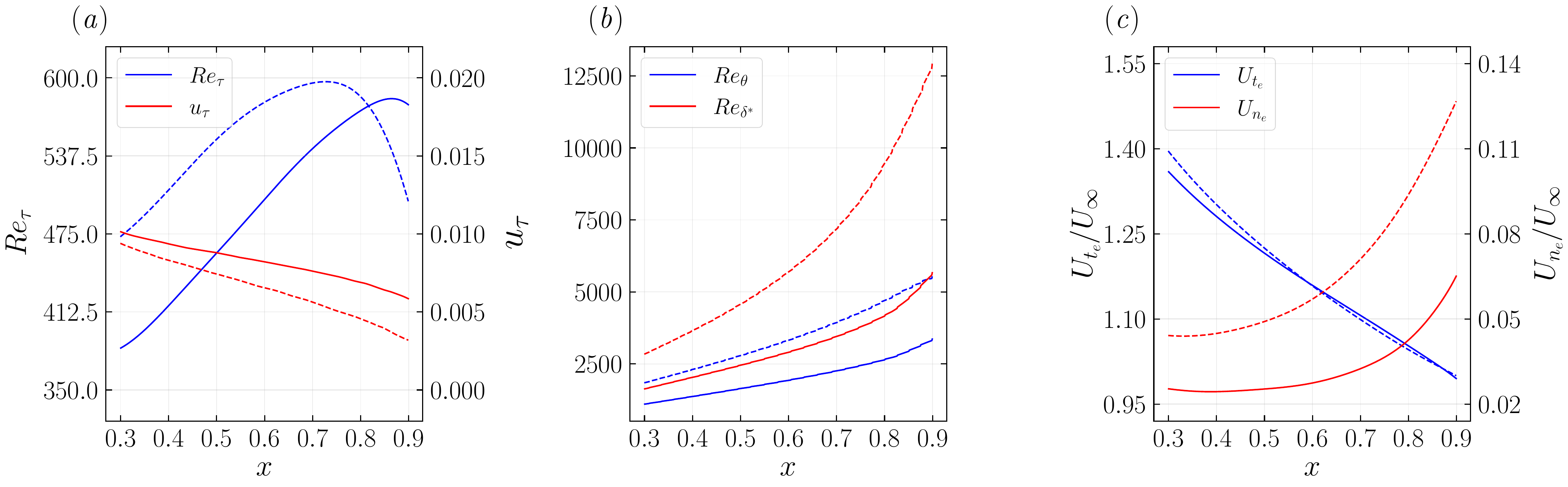}}
  \caption{Chordwise evolution of (\textit{a}) friction Reynolds number 
  $Re_\tau$ and friction velocity 
  $u_\tau$, (\textit{b}) Reynolds numbers based on the momentum thickness $Re_\theta$ and displacement thickness $Re_{\delta^*}$, and (\textit{c}) normalized tangential $U_{t_e}/U_\infty$ and wall-normal $U_{n_e}/U_\infty$ mean velocities along the edge of the boundary layer on the suction side of the airfoil for $9$ deg. (solid lines) and $12$ deg. (dashed lines).}
\label{fig:Re_x}
\end{figure}

The local values of the friction Reynolds number $Re_{\tau}=u_{\tau}\delta/\nu$, momentum-thickness Reynolds number $Re_{\theta}=U_{t_e}\theta/\nu$, and displacement-thickness Reynolds number $Re_{\delta^*}=U_{t_e}\delta^*/\nu$ are presented in figure \ref{fig:Re_x}. Here, $u_{\tau}$ is the friction velocity and $U_{t_e}$ is the tangential mean velocity at the boundary layer edge. The spatial variations of $Re_\tau$ and $u_\tau$ are presented in figure \ref{fig:Re_x}(\textit{a}). As can be observed, the friction Reynolds number increases along the chord until reaching maximum values of $Re_\tau = 583$ at $x = 0.86$ for the $9$ deg. case, and $Re_\tau = 597$ at $x = 0.73$ for $12$ deg. Downstream of these positions, both flows exhibit a decreasing behavior for $Re_\tau$ related to a steeper reduction in the friction velocity compared to the rise in the boundary layer thickness, both caused by the increasing APGs. A similar effect is observed by \citet{vinuesa2017} and \citet{tanarro2020} for a NACA4412 airfoil. 
Figure \ref{fig:Re_x}(\textit{b}) presents the distribution of $Re_\theta$ and $Re_{\delta^*}$ for both simulations. In their calculation, the boundary layer integral parameters $\theta$ and $\delta^*$ display an increasing behavior, while the streamwise mean velocity at the edge of the boundary layer $U_{t_e}$ decays, as shown in figure \ref{fig:Re_x}(\textit{c}). However, the growth of the integral quantities is comparatively larger than the reduction observed for the mean velocity. In particular, a steeper rise in $Re_{\delta^*}$ is observed for the $12$ deg. case in accordance with the results from figure \ref{fig:int_quant}. 
\citet{vinuesa2018} attributed this effect to the boundary layer thickening experienced due to the APG, and \citet{tanarro2020} reported a similar observation being related to the increase in the mean streamwise velocity deficit of the boundary layer due to the APG. Figure \ref{fig:Re_x}(\textit{c}) also displays the wall-normal mean velocity component at the edge of the boundary layer normalized by the freestream velocity $U_{n_e}/U_\infty$. As can be seen, this velocity increases with the APG, promoting an enhancement of the wall-normal convection, a result that is also observed by \citet{vinuesa2018}.

\subsection{Mean velocity profiles}\label{sec:mean_velocity}

\begin{figure}
  \centerline{\includegraphics[width=0.9\textwidth]{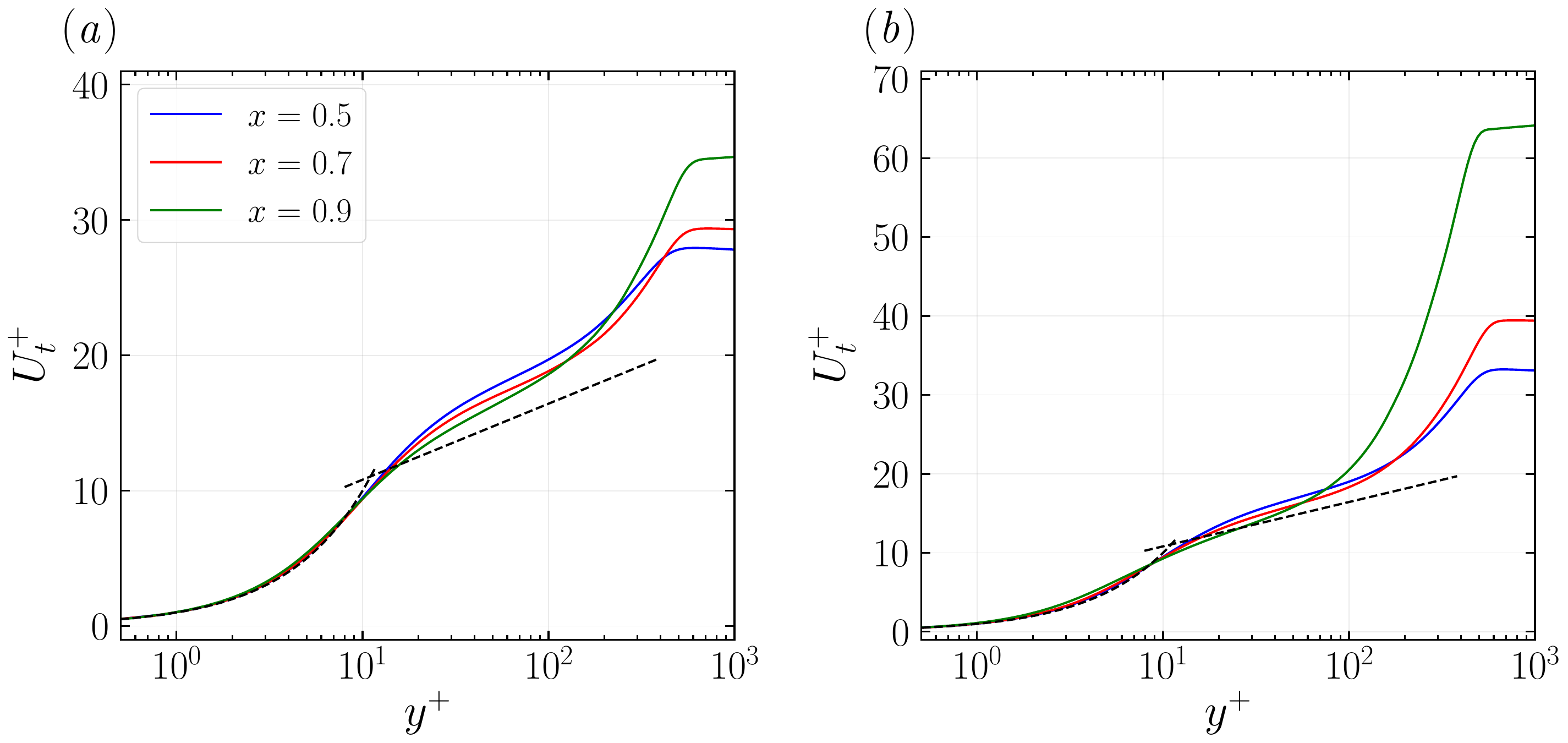}}
  \caption{Inner-scaled mean tangential velocity profiles at different chord locations for  (\textit{a}) $9$ and (\textit{b}) $12$ deg. angle of attack. Black dashed lines represent the law of the wall for the viscous sub-layer and log-layer, in which for the latter $\kappa = 0.41$ and $B = 5.2$.}
\label{fig:umais}
\end{figure}

The mean tangential velocity profiles are evaluated on the suction side at different chord positions being $x = 0.5$, $0.7$ and $0.9$. The respective values of the Clauser parameter for these positions are $\beta = 1.8$, $3.1$ and $9.4$, for 9 deg, and $\beta = 5.2$, $12.0$ and $46.2$, for 12 deg. These positions are chosen to assess different APG conditions, ranging from mild to moderate and strong APGs, depending on the angle of attack. Figure \ref{fig:umais} presents the inner-scaled mean velocity profiles as a function of the wall-normal distance (in wall units) for both simulations. One can observe the more prominent potential flow region of the scaled profiles with the increase in $\beta$, as also observed by \citet{samuel1974} and \citet{spalart1993}. 
For the case of $9$ deg. this effect is milder, however, for $12$ deg. the differences are significant, especially when comparing the positions $x = 0.5$ and $0.9$. This increase in the magnitude of the inner-scaled velocity is associated with the fact that the wall shear stress is reduced by the thickening of the boundary layer, i.e. the friction velocity is also reduced.

Another observation that can be drawn from figure \ref{fig:umais} is the deviation from the standard logarithmic law for the various APGs analyzed, which demonstrates that it may not be valid for some APG flows, as suggested by \citet{lee2008}. Moreover, the log layer of the profiles present a downward vertical shift for increasing APGs, a trend that was first observed by \citet{spalart1993}. The length of this region (in terms of wall units) is also reduced for higher APGs, as observed by \citet{monty2011}. In addition, as the APG increases, a larger slope is observed in the outer region, between the log-layer and the external potential flow region \citep{spalart1993,vinuesa2017}. Both previous effects can be seen more clearly for the 12 deg. incidence. This latter behavior is related to the APG effect on the momentum transfer across the boundary layer \citep{vinuesa2017}. 

\begin{figure}
  \centerline{\includegraphics[width=1\textwidth]{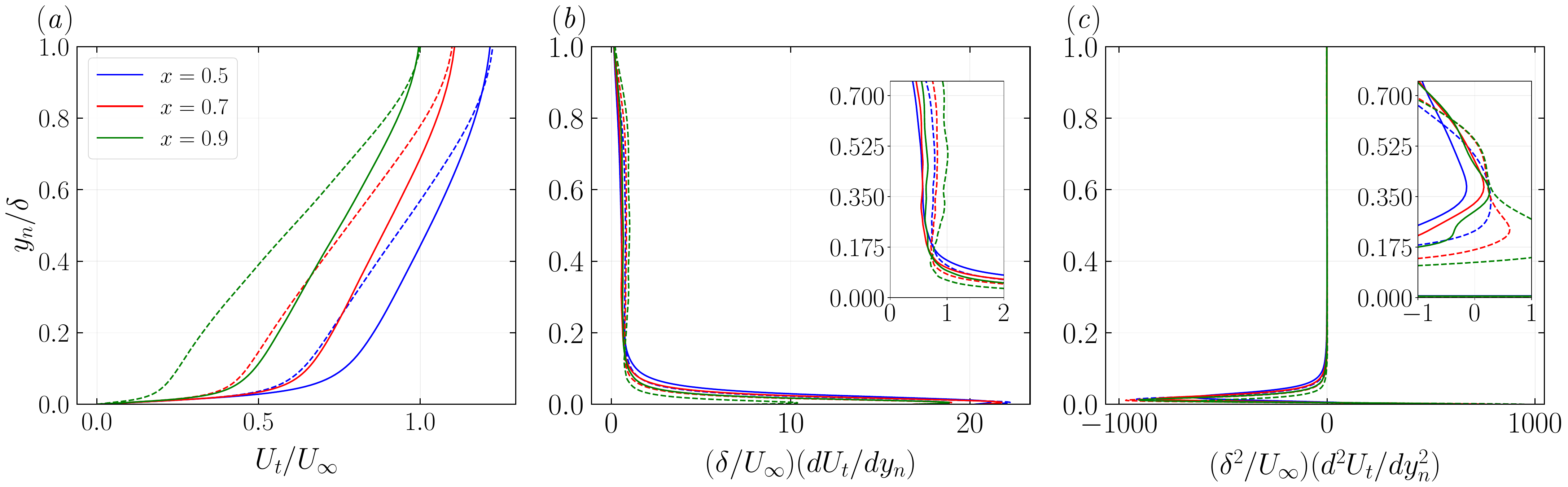}}
  \caption{Analysis of mean tangential velocity profiles at different chord positions: (\textit{a}) profiles scaled by the freestream velocity, (\textit{b}) first derivative, and (\textit{c}) second derivative of velocity profiles. Results are shown for $9$ (solid line) and $12$ deg. (dashed line) angle of attack.}
\label{fig:dudn}
\end{figure}

Figure \ref{fig:dudn}(\textit{a}) presents the mean tangential velocity profiles scaled by the freestream velocity as a function of the wall-normal distance $y_n$ scaled by the boundary layer thickness at different chord positions for both simulations. It can be observed that stronger APGs decelerate the flow near the wall. This effect changes the shape of the profiles, leading to the occurrence of inflection points which are observed for flows under strong APGs \citep{song2000}. In this context, an analysis of the flow stability properties can be conducted initially based on the Rayleigh criterion \citep{rayleigh1880}, which states that the existence of an inflection point is a necessary, but not sufficient condition for inviscid instability. Hence, figures \ref{fig:dudn}(\textit{b}) and \ref{fig:dudn}(\textit{c}) present the first and second derivatives of the velocity profiles, which allow a better identification of the inflection points and their characterization.
The plots of the second derivatives $U''_t = d^2U_t/dy_n^2$ (figure \ref{fig:dudn}(\textit{c})) show that all velocity profiles exhibit inflection points very close to the wall. Here, primes denote the differentiation with respect to the wall-normal coordinate. In this case, the values of $U''_t$ are positive at the wall, but rapidly decay to negative values. 
Other inflection points also appear in regions further away from the wall, except for the profile at $x = 0.5$ of the $9$ deg. angle of attack, which has a mild APG. These additional inflection points can be observed in the inset of figure \ref{fig:dudn}(\textit{c}). From this detail view, one can see that as the APG increases, the second inflection point moves towards the wall, while the third one moves away from it.

A stability analysis based on the inflection points can be further conducted from the perspective of the Fj{\o}rtoft criterion \citep{fjortoft1950}, which states that the necessary condition for inviscid instability is that $U''_t(U_t - U_{t_I}) < 0$ somewhere in the flow. This implies that the inflection point has to be a maximum of the mean vorticity away from the wall \citep{schmid2002}. Here, $U_{t_I}$ is the value of $U_t$ at the inflection point. 
Analyzing such points for both simulations and chord positions, it can be noted that, for the region just below the first inflection point, $(U_t - U_{t_I}) < 0$, while $U''_t > 0$, resulting in $U_t''(U_t - U_{t_I}) < 0$. In the region just above the first inflection point, $(U_t - U_{t_I}) > 0$ while $U''_t < 0$, which results in $U''_t (U_t - U_{t_I}) < 0$. Thus, this point satisfies the necessary and sufficient condition for inviscid instability. In the region just bellow the second inflection point $(U_t - U_{t_I}) < 0$ and $U''_t < 0$, which results in $U''(U_t - U_{t_I}) > 0$. On the other hand, in the region just above it $(U_t - U_{t_I}) > 0$ and $U''_t > 0$, also resulting in $U''_t (U_t - U_{t_I}) > 0$. Hence, this point does not satisfy the criteria for inviscid instability. Finally, analyzing the region just below the third inflection point, it is found that $(U_t - U_{t_I}) < 0$ while $U''_t > 0$, resulting in $U''_t (U_t - U_{t_I}) < 0$. The region just above this point has $(U_t - U_{t_I}) > 0$ and $U''_t < 0$, which results in $U''_t (U_t - U_{t_I}) < 0$. Therefore, the third inflection point satisfies both the \citet{rayleigh1880} and the \citet{fjortoft1950} criteria for inviscid instability.

Here, we examine the unstable inflection points of the velocity profiles with more scrutiny. The analysis of the first point is performed recalling the ideas from \citet{marquillie2011}. These authors performed a linear stability analysis of a TBL channel flow subjected to an APG by superimposing near-wall streaks in the mean velocity profile. They concluded that the streaks are responsible for instability modes, and discussed that such instability is related to the peak of turbulent kinetic energy which appears near the wall. Here, the first inflection point presented in figure \ref{fig:dudn} is related to the action of streaks lying in a region close to the wall, where viscous effects are important to the TBL. On the other hand, the third inflection point appears in a region where the viscous effects are not as important, indicating the presence of an inviscid instability mechanism that may originate an embedded shear layer to the TBL, as also observed by \citet{schatzman2017}.

\subsection{Reynolds stresses}

\begin{figure}
  \centerline{\includegraphics[width=0.9\textwidth]{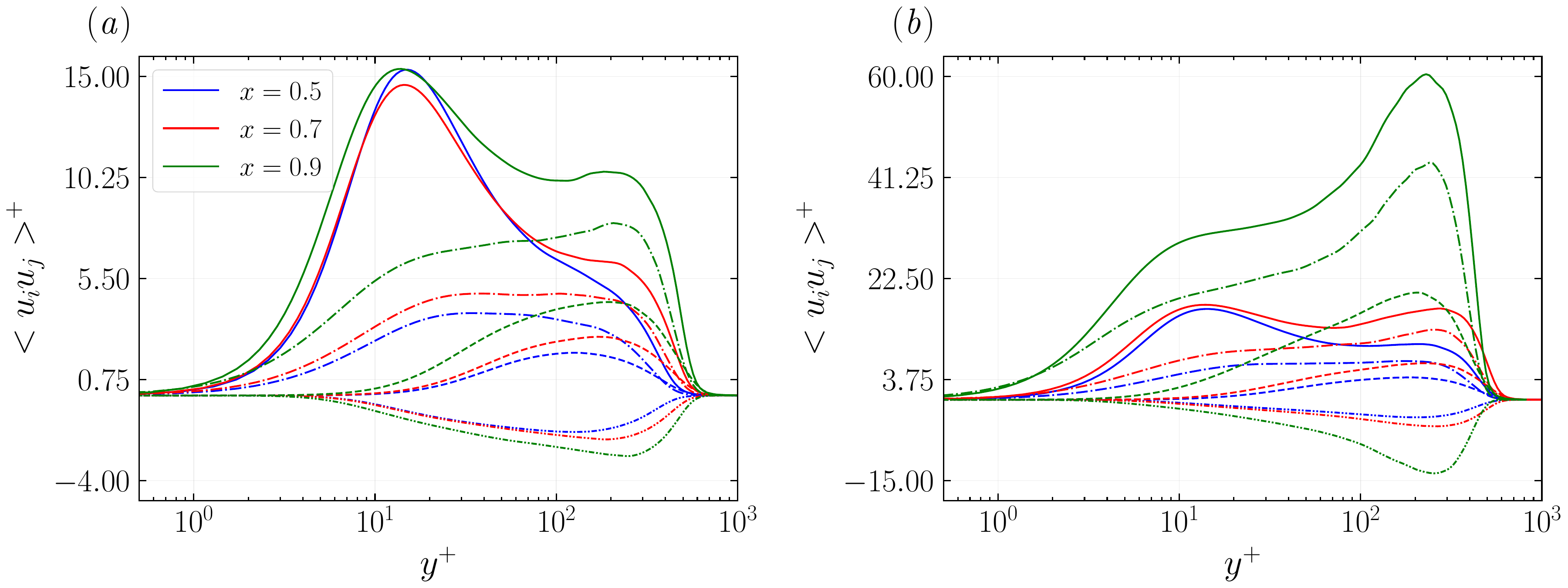}}
  \caption{Reynolds stresses at different chord locations for  (\textit{a}) $9$ and (\textit{b}) $12$ deg. angle of attack. The line styles indicate: (\protect\solidline) $\left< u_tu_t \right>$, (\protect\dashedline) $\left< u_nu_n \right>$, (\protect\dashdotted) $\left< ww \right>$ and, (\protect\densdashdot) $\left< u_tu_n \right>$.}
\label{fig:reynolds}
\end{figure}

Additional understanding of the APG effects for the present TBLs can be acquired through inspection of the Reynolds stresses. Figure \ref{fig:reynolds} presents the normal components of the Reynolds stress tensor (tangential $\langle u_t u_t \rangle$, wall-normal $\langle u_n u_n \rangle$, and spanwise $\langle w w \rangle$), as well as the shear stress $\langle u_t u_n \rangle$, scaled by the friction velocity $u_{\tau}^2$ at the same chord positions analyzed in the previous section. 
It can be observed that for all chord locations of the $9$ deg. case, shown in figure \ref{fig:reynolds}(\textit{a}), the first peaks of the tangential Reynolds stresses have similar magnitudes despite the different APGs. A similar observation can be made from figure \ref{fig:reynolds}(\textit{b}) for $x = 0.5$ and $0.7$ of the $12$ deg. case, and these results are in agreement with the literature \citep{monty2011}. However, this particular scaling does not fully collapse the profiles in the near-wall region. Moreover, for the $12$ deg. setup at $x=0.9$, the magnitude of the APG (in terms of $\beta$) is around 4.9 times that computed for the same position of the $9$ deg. incidence configuration, as shown in figure \ref{fig:int_quant}(\textit{b}). For this case, the inner peak of $\langle u_t u_t\rangle^+ $ is more pronounced, reaching a value two times higher than those computed for other chord locations.

The analysis of the outer region shows that the impact of the APG is even more remarkable. The tangential component of the Reynolds stresses start developing a secondary peak as the APG increases. This behavior can be first noticed for $x = 0.7$ and $0.9$ of the $9$ deg. setup. A mild APG is encountered for the former position, leading to a small plateau, and a moderate APG is found for the latter, resulting in a second peak. For the $12$ deg. angle of attack configuration, the APGs are stronger and the previous trend is even more noticeable. For example, the secondary peak of $\langle u_t u_t \rangle$ reaches a higher value than that obtained for the inner peak at $x = 0.9$. This phenomenon is explained by the action of the APG on the larger, most energetic scales of the boundary layer, which results in a higher turbulence intensity in the outer region \citep{monty2011,harun2013,vinuesa2017}.

For ZPG-TBLs, the spanwise Reynolds stress $\langle w w \rangle^+$ depicts a growth near the wall and a smooth decay towards the edge of the boundary layer \citep{pope2000,schiavo2015}. However, for the present APGs, similar features observed for $\langle u_t u_t \rangle^+$ appear for $\langle w w \rangle^+$ and, under strong APGs, the second peak of $\langle w w \rangle^+$ is more pronounced than the first one.
An inspection of the wall-normal component of the Reynolds stresses $\langle u_n u_n \rangle^+$ shows that the outer region is also energized by wall-normal velocity fluctuations due to rises in the APG. This is verified for both angles of attack investigated and, as discussed by \citet{vinuesa2018} and also shown in figure \ref{fig:Re_x}(\textit{c}), the APG increases the mean wall-normal velocity, contributing to the thickening of the boundary layer. In the same fashion, the wall-normal velocity fluctuations are also enhanced in the outer layer, leading to the peaks observed in figure \ref{fig:reynolds} for $y^+ > 10^2$. 
The same conclusions can be drawn from the inspection of the shear stress component $\langle u_t u_n \rangle^+$, which is more significant in the outer region of the boundary layer than near the wall. The figure shows that this component increases with the APG, as also reported by \citet{skaare1994}, and this behavior leads to different mechanisms of momentum distribution along the boundary layer caused by the APG \citep{vinuesa2017}. In summary, all components of the Reynolds stress tensor are influenced by the presence of the APG due to the fact that both the large and small scales are energized in the outer layer \citep{monty2011,tanarro2020}, and this results in an increase of the turbulence intensity in this region.

\subsection{Turbulent kinetic energy budgets}

\begin{figure}
  \centerline{\includegraphics[width=1\textwidth]{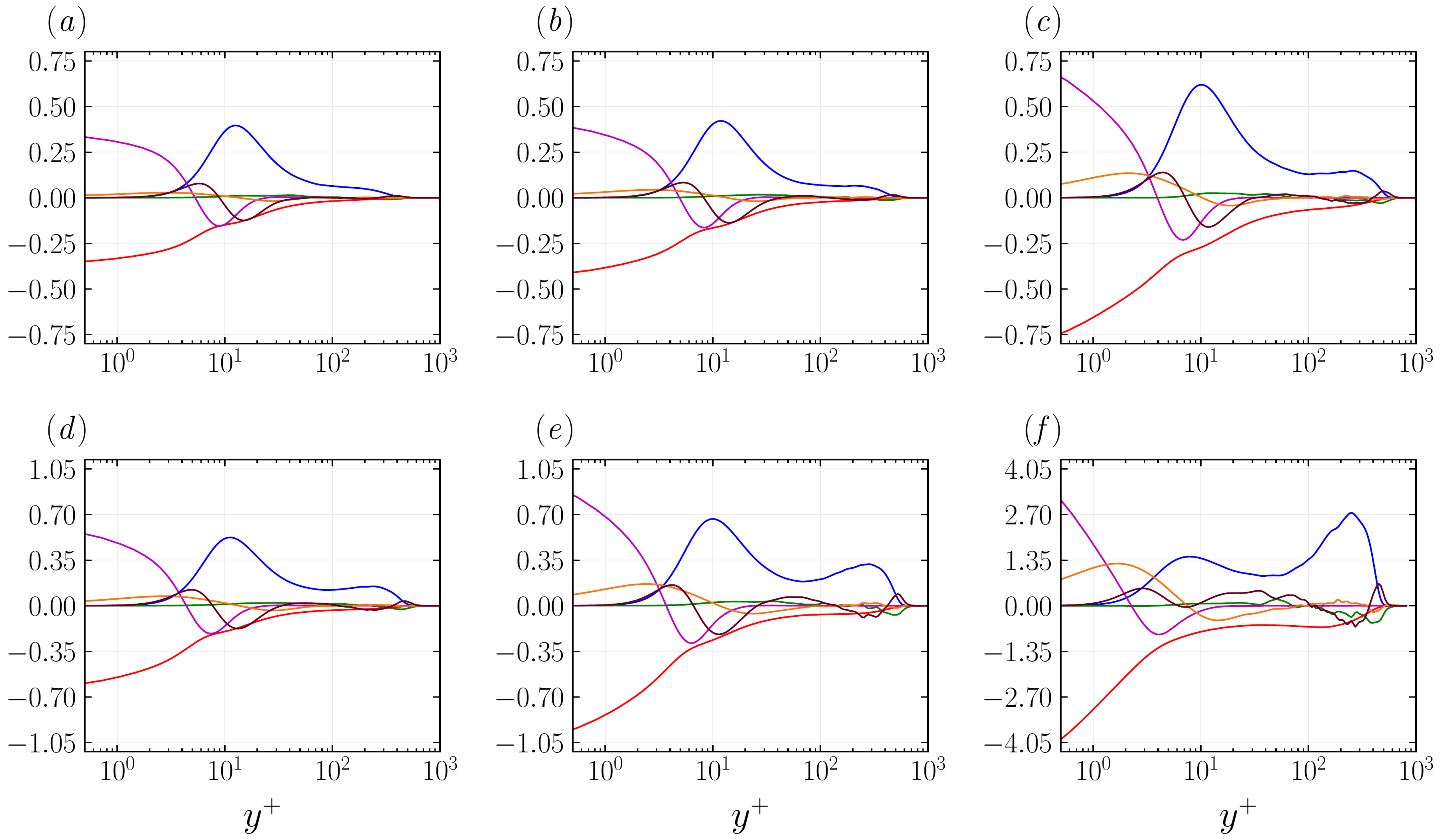}}
  \caption{Turbulent kinetic energy budgets scaled by $u_{\tau}^4/\nu$. Plots on the top and bottom rows present results for $9$ and $12$ deg. angle of attack, respectively. 
  Profiles are computed at (\textit{a}, \textit{d}) $x=0.5$, (\textit{b}, \textit{e}) $x=0.7$ and (\textit{c}, \textit{f}) $x=0.9$. The colors indicate: (\textcolor{blue}{$\raisebox{0.6mm}{\rule{3mm}{0.5mm}}$}) Production, (\textcolor{red}{$\raisebox{0.6mm}{\rule{3mm}{0.5mm}}$}) Pseudo-dissipation, (\textcolor{Green}{$\raisebox{0.6mm}{\rule{3mm}{0.5mm}}$}) Advection, (\textcolor{DarkOrchid}{$\raisebox{0.6mm}{\rule{3mm}{0.5mm}}$}) Viscous diffusion, (\textcolor{Brown}{$\raisebox{0.6mm}{\rule{3mm}{0.5mm}}$}) Turbulent transport, and (\textcolor{orange}{$\raisebox{0.6mm}{\rule{3mm}{0.5mm}}$}) Pressure diffusion.}
\label{fig:tke}
\end{figure}

To further understand the APG effects on the TBLs investigated, the turbulent kinetic energy (TKE) budget is evaluated. Figure \ref{fig:tke} presents the TKE scaled by $u_{\tau}^4/\nu$. The top row shows results for the $9$ deg. simulation while the bottom one presents those for $12$ deg. at chord positions $x = 0.5$, $0.7$, and $0.9$. The first term that presents a noticeable effect of the APG is the production. For both simulations, the inner layer peak is increased with the APG. One can also observe that as the APG increases, the first peak starts to shift towards the wall. In the $9$ deg. simulation, at $x = 0.5$ (figure \ref{fig:tke}(\textit{a})), the first peak is located at $y^+ \approx 12.5$, while at $x = 0.9$ (figure \ref{fig:tke}(\textit{c})), it is located at $y^+ \approx 10.0$. The same behavior is observed for the $12$ deg. angle of attack, where at $x = 0.5$ (figure \ref{fig:tke}(\textit{d})), the first peak is located at $y^+ \approx 11.0$, while at $x = 0.9$ (figure \ref{fig:tke}(\textit{f})), it is located at $y^+ \approx 8.0$. Note that this also applies when comparing the same chord positions for the two simulations, where the APGs are stronger for the higher angle of attack. The production term also presents an increase in the outer region forming a plateau for mild and moderate APGs and a second peak for strong APGs. The second peak reaches even larger values than the inner peak at $x = 0.9$ for $12$ deg. angle of attack (figure \ref{fig:tke}(\textit{f})). This is also observed by \citet{skaare1994} who related such effect to the increase in the turbulent shear stresses in the outer region of the boundary layer, which can be also noticed from figure \ref{fig:reynolds}. Moreover, the pronounced second peak in the outer region presented in the production term is an evidence that the APG energizes the large-scale motions of the flow as observed by \citet{harun2013}.

Significant effects are also observed in the pseudo-dissipation term for all cases analyzed. The magnitude of this term is increased throughout the entire boundary layer, with the amplifications being more evident at the wall, where dissipation is balanced by the pressure and viscous diffusion terms. These observations are in line with what was presented by \citet{skaare1994}, who demonstrated that the  diffusion terms become more important with the increase in the APG. The magnitude of the viscous diffusion also increases with the APG and its negative counterpart acts with the dissipation term to balance the first peak of production. The pressure diffusion increases in the region close to the wall  balancing the pseudo-dissipation for high APGs as shown in figure \ref{fig:tke}(\textit{f}). Finally, near the edge of the boundary layer, a negative peak of the turbulent transport term appears in the outer region for high APGs. This term balances the second production peak in conjunction with the pseudo-dissipation and advection terms. At the outermost part of the boundary layer, the advection and turbulent transport terms present negative and positive peaks, respectively. These are related to the interaction between the structures at the edge of the boundary layer and the freestream flow \citep{vinuesa2017}.

\subsection{Analysis of turbulence production}

The TKE production term is directly related to the energization of the TBL and, therefore, it is important to evaluate its components. Due to the spanwise periodicity of the simulations, the production is statistically two-dimensional, i.e., the flow homogeneity leads to a zero mean spanwise velocity and its derivatives. With these considerations, the TKE production $P_k$ is written as:
\begin{equation}\label{eqn:production}
    P_k = \underbrace{-\left<u_t u_t \right>\frac{\partial U_t}{\partial x_t}}_{P_1} \underbrace{-\left<u_t u_n \right>\frac{\partial U_t}{\partial y_n}}_{P_2} \underbrace{-\left<u_t u_n \right>\frac{\partial U_n}{\partial x_t}}_{P_3} \underbrace{-\left<u_n u_n \right>\frac{\partial U_n}{\partial y_n}}_{P_4} \mbox{ .}
\end{equation}

\begin{figure}
  \centerline{\includegraphics[width=1\textwidth]{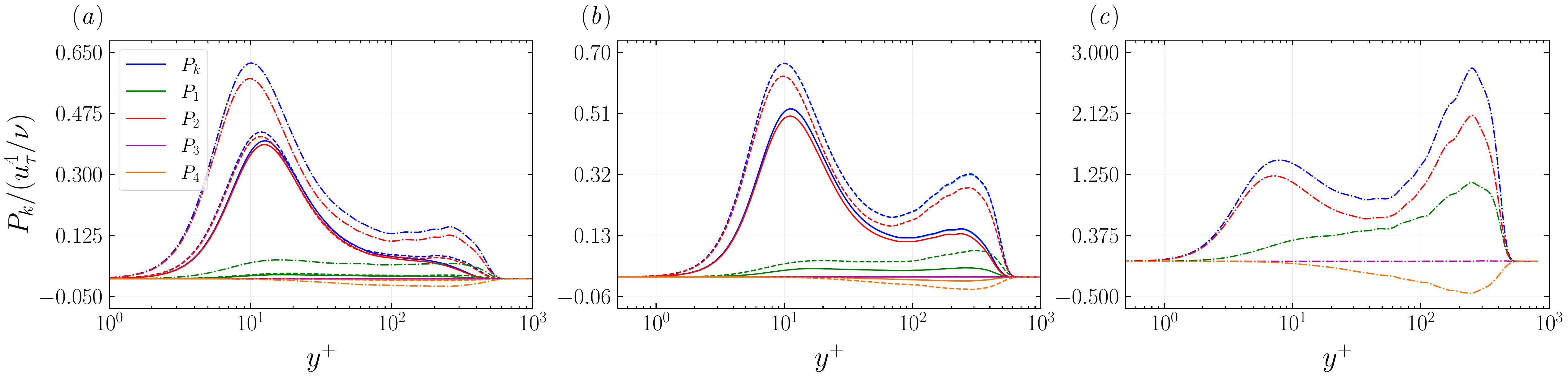}}
  \caption{Production terms for angles of attack (\textit{a}) $9$ deg, (\textit{b}) $12$ deg. at postions $x = 0.5$ and $0.7$, and (\textit{c}) $12$ deg. at $x = 0.9$. The line styles indicate: (\protect\solidline) $x = 0.5$, (\protect\dashedline) $x = 0.7$, (\protect\dashdotted) $x = 0.9$.}
\label{fig:prod_terms}
\end{figure}

Figure \ref{fig:prod_terms} presents the production along with its components for both simulations at positions $x = 0.5$, $0.7$ and $0.9$. All the components are scaled by $u_{\tau}^4/\nu$. Figure \ref{fig:prod_terms}(\textit{a}) shows the results for angle of attack 9 deg., while figures \ref{fig:prod_terms}(\textit{b}) and (\textit{c}) depict the results for 12 deg. The middle figure brings the results for $x = 0.5$ and 0.7, while the right one displays the production for $x = 0.9$. This choice is for better visualization purposes due to the larger values of production observed in the most downstream position of the 12 deg. case.
One can observe that the main contribution to $P_k$ remains the $P_2$ term, which is related to the interaction of the Reynolds shear stress with the mean shear.
The influence of these two parameters can be seen through figures \ref{fig:dudn}(\textit{b}) and \ref{fig:reynolds}. Note that close to the wall the dominant term in $P_2$ is the mean shear while the Reynolds shear stress remains small. On the other hand, in the outer layer, the dominant parameter becomes the Reynolds shear stress since the mean shear decreases rapidly. It is interesting to note that the second inflection point presented in figure \ref{fig:dudn}(\textit{c}) corresponds to the position where $P_k$ starts to increase in the outer layer. Moreover, similarly to the observation from \citet{schatzman2017}, the third inflection point matches the position of the maximum value of the $P_2$ term and the negative peak value of the Reynolds shear stress.

For strong APGs, significant contributions to the TKE production are also made by the $P_1$ and $P_4$ terms. In the near-wall region, the $P_1$ has a smaller impact in $P_k$ since the streamwise derivative of $U_t$ is low. However, both the magnitude of $\partial U_t / \partial x_t$ as well as $\left< u_t u_t \right>$ increase in the outer region due to the APG. Hence, this component of the production term also contributes to increasing the secondary peak of $P_k$ observed in figure \ref{fig:prod_terms}. On the other hand, $P_4$ contributes to a reduction of the total TKE production mainly in the outer region. This effect is more evident as the APG increases, since it impacts both $\partial U_n / \partial y_n$ and $\left< u_n u_n \right>$.

\begin{figure}
  \centerline{\includegraphics[width=1\textwidth]{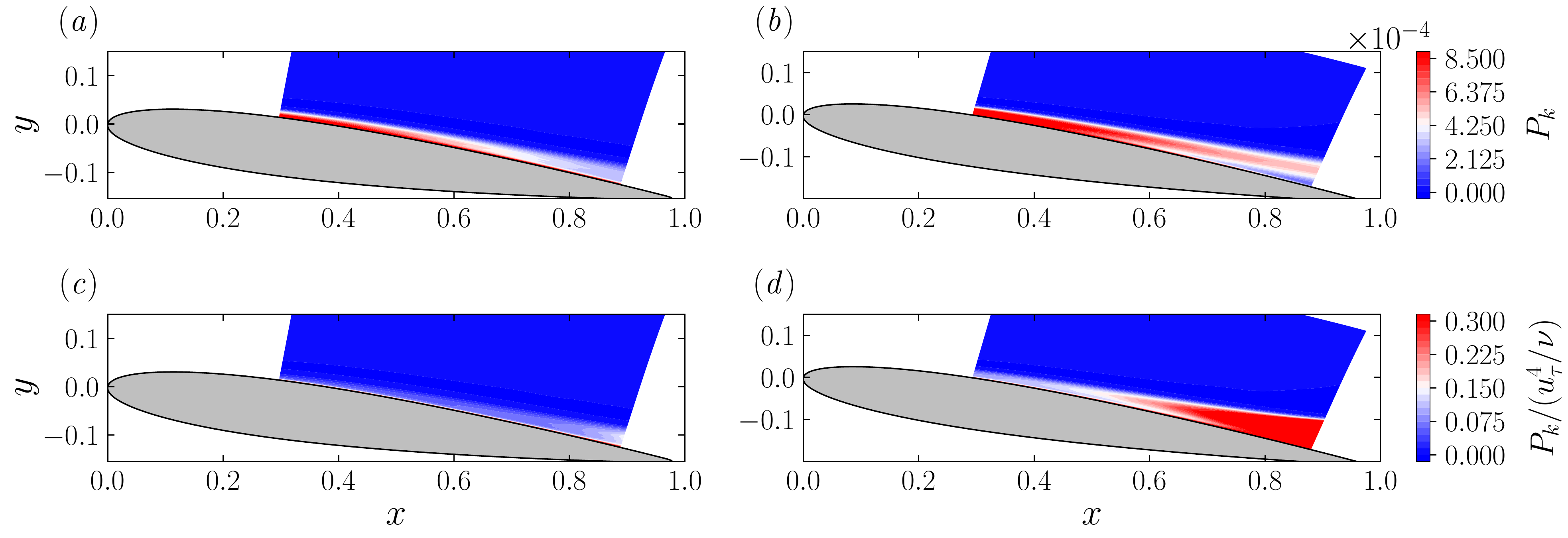}}
  \caption{Spatial distribution of TKE production. The left and right columns present results for the $9$ and 12 deg. simulations, respectively, for (\textit{a}, \textit{b}) production without scaling, and (\textit{c}, \textit{d}) production scaled by $u_{\tau}^4/\nu$.}
\label{fig:prod_maps}
\end{figure}

As can be observed in figures \ref{fig:reynolds} and \ref{fig:tke}, the standard inner scaling does not collapse the Reynolds stress and TKE mean profiles and, hence, it may not be appropriate for APG flows. This issue was discussed by \citet{maciel2018}, who analyzed scaling effects. Although this is not the main goal of this study, the following observation can be drawn from our results and the impact of the inner scaling is demonstrated for the production term. In figure \ref{fig:prod_maps}, the top row presents the spatial distribution of the production term without scaling, while the bottom row shows the production scaled by $u_{\tau}^4/\nu$. Results are shown for 9 and 12 deg. on the left and right columns, respectively.
The top figures show that the production is higher at upstream positions along the chord, in the near-wall region, and this is due to the higher mean shear effects. \citet{skaare1994} observed that the turbulence intensity tends to decrease in the inner region as the APG increases and this is also demonstrated in figures \ref{fig:prod_maps}(\textit{a}) and (\textit{b}). 

As can be seen, the production is reduced downstream and displays two peaks, being one very close to the wall and another in the outer layer. This separation becomes more evident for the airfoil at 12 deg. angle of attack in figure \ref{fig:prod_maps}(\textit{b}), which shows the delelopment of a shear layer away from the wall. The physical features just described are not properly captured by the inner scaling as can be observed in figures \ref{fig:prod_maps}(\textit{c}) and (\textit{d}). This is due to the reduction in the friction velocity $u_{\tau}$ with the increase in the APG, as shown in figure \ref{fig:Re_x}. 
To remedy this issue, more appropriate outer scale variables have been used by some authors \citep{maciel2018, schatzman2017, wei2023} who studied APG-TBLs. In these cases, shear-layer variables have been used, providing better scalings.

\subsection{Characterization of anisotropy state}

In order to further assess the APG effects along the present boundary layers, the state of anisotropy of the Reynolds stresses is characterized under the perspective of the Lumley triangle \citep{choi2001}. The analysis is presented through the trajectories of the invariants of the normalized anisotropy tensor $b_{ij} =\left< u_i^\prime u_j^\prime \right>/\left< u_k^\prime u_k^\prime \right> - \delta_{ij}/3$ in the Lumley triangle in the wall-normal direction \citep{schiavo2017}. 
Three independent invariants of $b_{ij}$ can be described as:
\refstepcounter{equation}
$$
  I = b_{ii}, \quad
  II = -\frac{b_{ij}b_{ji}}{2}, \quad
  III = \frac{b_{ij}b_{jk}b_{ki}}{3} \mbox{ .}
 \eqno{(\theequation)}
 \label{eqn:inv}
$$

One should note that $I$ is the trace of the anisotropy tensor $b_{ij}$ and, therefore, it is identically zero. Consequently, the anisotropy is characterized by the second and third invariants. In order to evaluate the nonlinear behavior in the return to isotropy, \citet{choi2001} introduced the variables $\xi = \left( \frac{III}{2} \right)^{1/3}$ and $\eta = \left( -\frac{II}{3} \right)^{1/2}$, which allow a characterization of the anisotropy state by plotting the $\xi-\eta$ values in the Lumley triangle.

\begin{figure}
  \centerline{\includegraphics[width=1\textwidth]{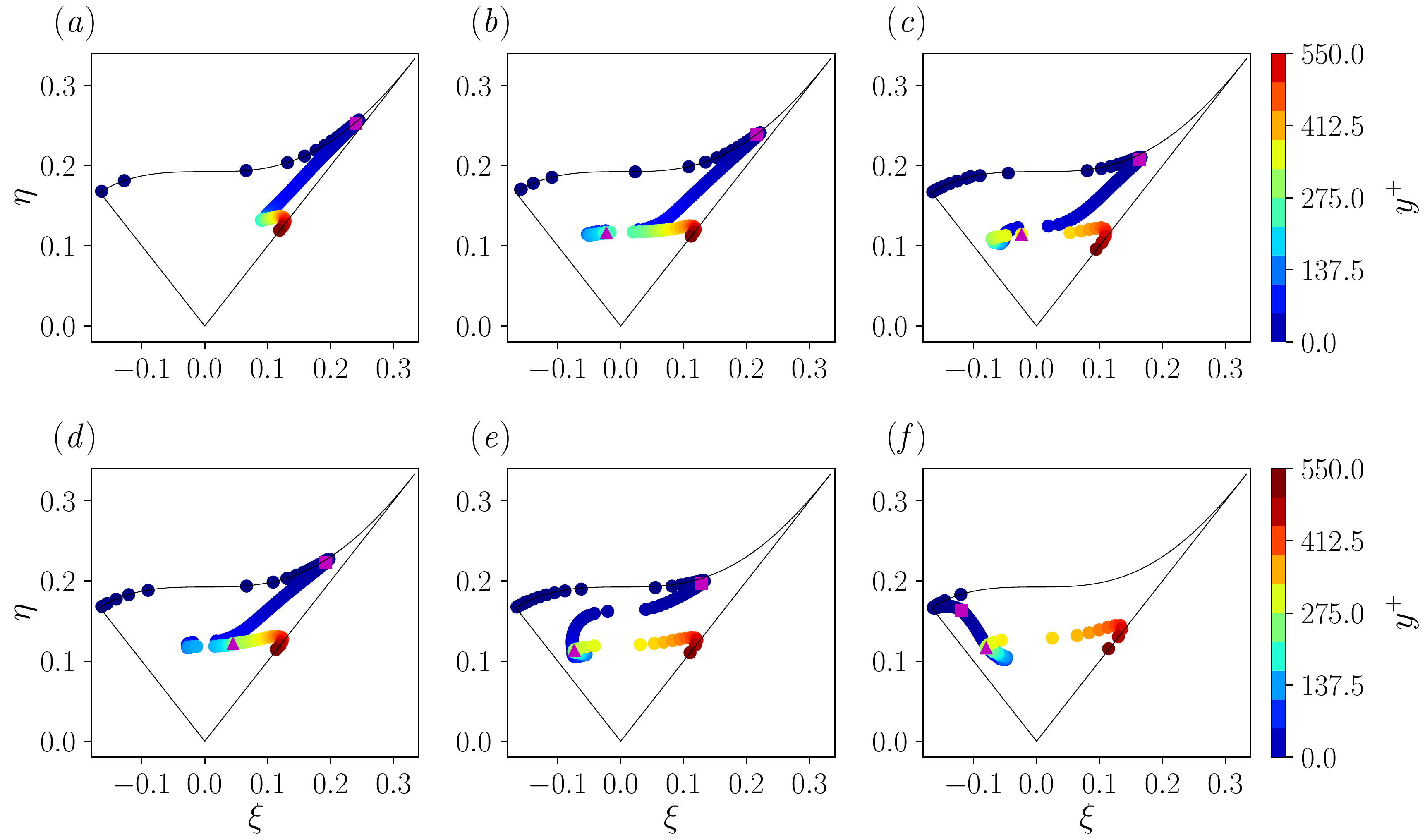}}
  \caption{Lumley triangle evaluated at positions (\textit{a}, \textit{d}) $x = 0.5$, (\textit{d}, \textit{e}) $x = 0.7$, and (\textit{c}, \textit{f}) $x = 0.9$. The figures on top and bottom rows correspond to the $9$ and 12 deg. simulations, respectively. The symbols $\blacksquare$ and $\blacktriangle$ represent the position of the first and second peaks of the TKE production, respectively.}
\label{fig:lumley}
\end{figure}

Figure \ref{fig:lumley} presents the Lumley triangles computed at $x = 0.5$, $0.7$, and $0.9$ for both simulations. In the triangles,  isotropic turbulence is represented at the origin of the $\xi-\eta$ coordinates. The left corner represents a 2D isotropic turbulence state, while the right corner represents 1D turbulence. The top curve that connects these two corners represents a 2D turbulence state. Along the line $\eta = -\xi$, the turbulence becomes axisymmetric so that one eigenvalue of the normalized anisotropy tensor is small while the others have comparable larger magnitudes. On the other hand, along the line $\eta = \xi$, the turbulence is also axisymmetric, but one eigenvalue is larger than the other two.

The colors in the trajectories of figure \ref{fig:lumley} represent the turbulence states as a function of the wall-normal distance in the boundary layer, and the symbols $\blacksquare$ and $\blacktriangle$ represent the position of the first and second peaks appearing in the Reynolds stress $\langle u_t u_t \rangle^+$ distribution 
and TKE production.
The plots in the top row of the figure show the results for the $9$ deg. simulation. The case of a mild APG is presented in figure \ref{fig:lumley}(\textit{a}), which is computed for $x = 0.5$. At the near-wall region, a 2D state is observed and, as $y^+$ increases, the turbulence approaches the 1D condition, due to near-wall streaks which lead to high values of $\langle u_t u_t \rangle^+$. Then, the trajectory follows the line of axisymmetric expansion ($\eta = \xi$). 
This behavior is typically observed for ZPG wall-bounded flows and for mild APGs \citep{pope2000,nogueira2021}. However, as can be seen from figures \ref{fig:lumley}(\textit{b}) and (\textit{c}), obtained for chord positions $x = 0.7$ and 0.9, respectively, the increase in the APG causes the turbulence states to move away from the 1D turbulence condition, as well as from the axisymmetric expansion line. For these cases, after the first peak of production, one can see that the turbulence state approaches that of the axisymmetric contraction line ($\eta=-\xi$) in the outer layer. For both cases, the turbulence state at the second peak of production is located near the center of the triangle, before the trajectory returns to the axisymmetric expansion condition.

The bottom row presents results for the $12$ deg. simulation and the trajectory for $x=0.5$, shown in figure 
\ref{fig:lumley}(\textit{d}) is similar to that of 9 deg. at $x=0.9$. However, the changes in the trajectories for $x=0.7$ and 0.9 are more pronounced compared with those computed for the $9$ deg. angle of attack case due to the stronger APGs. A strong APG ($\beta \approx 46.2$) is encountered at $x = 0.9$ for the $12$ deg. angle of attack case and its anisotropy state is presented in figure \ref{fig:lumley}(\textit{f}). This plot shows that, in the near-wall region, the trajectory never approaches the 1D turbulence state. Instead, it stays close to the 2D isotropic state even at the first peak of TKE production and, then, it follows the axisymmetric contraction region. After the second peak of production, it moves to the axisymmetric expansion line, as observed for all cases analyzed.

The present results show that, as the APG increases from mild to moderate values, the trajectories of the invariants are shifted towards the axisymmetric contraction condition. For such cases, the turbulence state resembles that of a shear-layer \citep{pope2000,biancofiore2014}. Moreover, under the strong APG condition computed at $x=0.9$ for $12$ deg, the trajectory of the invariants is similar to that presented by \citet{schiavo2017} for a separated boundary layer over a smooth bump profile. This may suggest that the present airfoil boundary layer is in the vicinity of separation.

\begin{figure}
  \centerline{\includegraphics[width=0.9\textwidth]{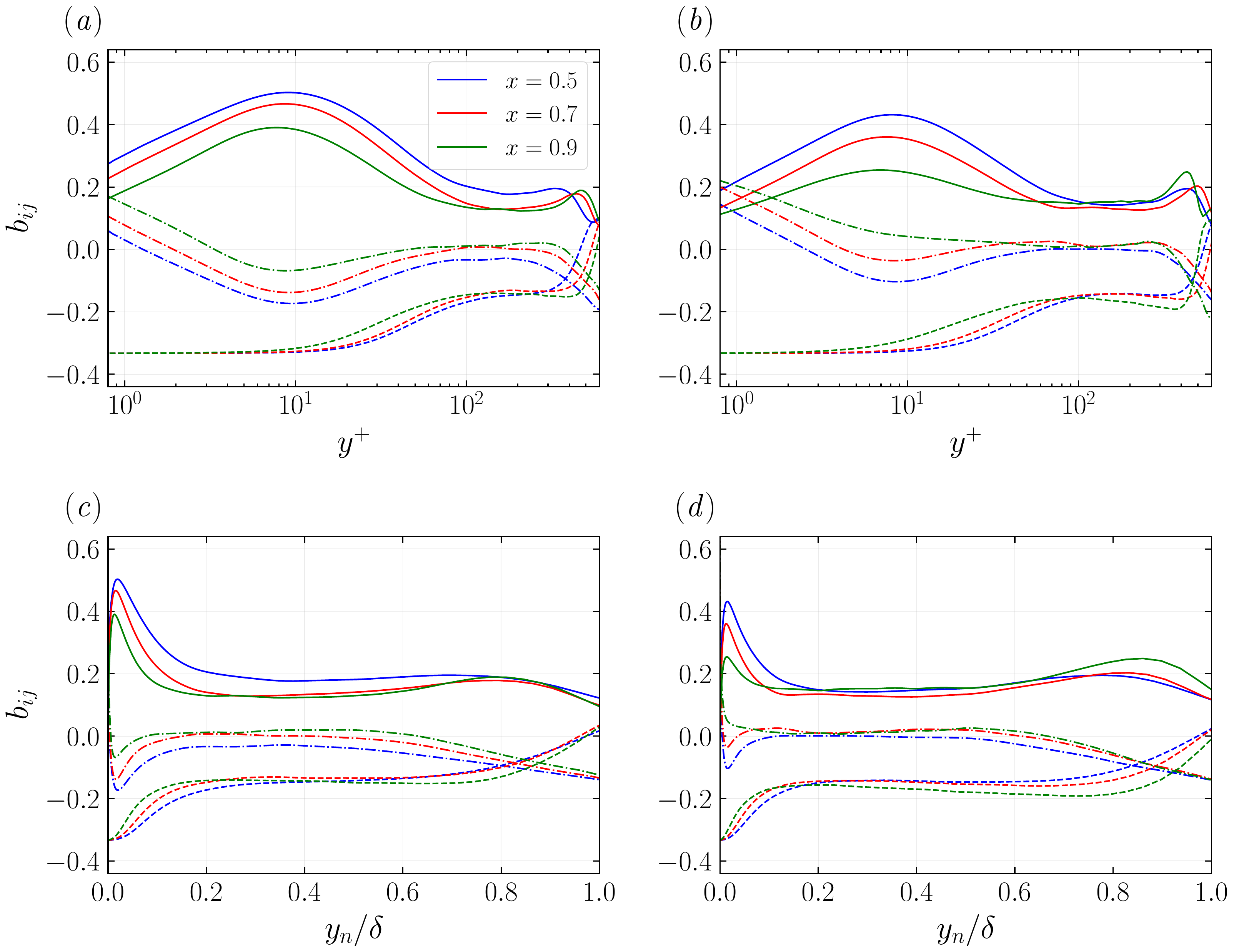}}
  \caption{Anisotropy tensor with (\protect\solidline) $b_{11}$, (\protect\dashedline) $b_{22}$ and, (\protect\dashdotted) $b_{33}$. The left and right columns present results for $9$ and 12 deg, respectively. The top and left rows present results normalized by the viscous length scale and the boundary layer thickness, respectively, allowing separate assessments of the near-wall region and the outer layer.}
\label{fig:betaij}
\end{figure}

Figure \ref{fig:betaij} displays the anisotropy tensor components for the cases presented in figure \ref{fig:lumley}, corroborating the previous analysis. 
The left and right columns present results for the $9$ and 12 deg. simulations, respectively. The top row presents the results in terms of wall units, while the bottom plots are shown  normalized by the boundary layer thickness. Hence, the figures allow a separate analysis of the turbulence states in the near-wall region and in the outer layer.
For both cases, the $b_{11}$ and $b_{33}$ components of the normalized anisotropy tensor dominate in the viscous sublayer, which characterize the 2D turbulence state observed in the Lumley triangle. As typically observed for TBLs, a peak of $b_{11}$ is observed in the buffer region at $y^+ \approx 10$. This is displayed by the $\blacksquare$ symbols in figure \ref{fig:lumley} and one can see from figure \ref{fig:betaij}  that, as the APG increases, the $b_{11}$ peaks have reduced values and move towards the wall. In the same context, the values of $b_{33}$ increase and this effect leads to the lack of a quasi-1D state for strong APGs, as shown in figure \ref{fig:lumley}(\textit{f}).

After the first peak, $b_{11}$ decreases and achieves a plateau followed by a small bump observed in the outer layer. At the same time, the growth of $b_{22}$ is noticed. The second peak of production is observed at $y^+ \approx 200$ and this is the region where $b_{11}$ and $b_{33}$ are larger than $b_{22}$, i.e., that of axisymmetric contraction in the Lumley triangles. This effect is more pronounced for stronger APGs. Right after the plateau, one can see the increase of $b_{11}$ and the decrease of $b_{33}$, which causes the turbulence state to move from the axisymmetric contraction state to the axisymmetric expansion one. Then, towards the edge of the boundary layer, the dominant components become $b_{11}$ and $b_{22}$, as can be seen from the bottom plots. This would bring the turbulence states back to the axisymmetric contraction region of the Lumley triangles, as discussed by \citet{biancofiore2014}, who studied shear turbulent layers. These latter states are not shown in figure \ref{fig:lumley} to improve the visualization.

\subsection{Inspection of coherent structures}

In order to gain further understanding of the physical processes in the inner and outer layers, we investigate the coherent motion of turbulent structures in the present APG-TBLs. For this, the proper orthogonal decomposition (POD) is employed. This method was first introduced by \citet{lumleyl1967} and it consists of constructing an optimal basis to extract modes optimizing the data variance. In the context of fluid mechanics, these optimal bases should represent coherent flow structures \citep{holmes2012}. The application of POD begins through the decomposition of a spatio-temporal field $q(\textbf{x},t)$ into a temporal mean $\langle q(\textbf{x})\rangle$ and its fluctuation $q'(\textbf{x},t)$. The latter is then decomposed into spatial $\phi_i(\textbf{x})$ and temporal modes $a_i(t)$ as:
\begin{equation}\label{eqn:q}
  q(\textbf{x},t) =  \langle q(\textbf{x})\rangle + q'(\textbf{x},t) = \langle q(\textbf{x})\rangle + \sum_{i=1}^{m} a_i(t)\phi_i(\textbf{x}) \mbox{ .}
\end{equation}

To compute the POD modes, we first organize the fluctuation fields in a matrix $\textbf{Q} \in \mathbb{R}^{n \times m}$ in which the columns represent the flow temporal evolution with $m$ time samples. The lines represent a stacked array with $n_p$ flow properties along the $n_g$ grid points where $n=n_p \times n_g$. Then, a temporal correlation matrix of the data, $\textbf{C} = \textbf{Q}^T\textbf{Q} \in \mathbb{R}^{m \times m}$, is computed using the snapshot method proposed by \citet{sirovich1987} since $n \gg m$ for the present simulations. 
Moreover, we employ the spectral POD (SPOD) method proposed by \citet{sieber2016} as
\begin{equation}\label{eqn:c}
  S_{i,j} =  \sum_{k=-N_f}^{N_f} g_{k}C_{i+k,j+k} \mbox{ ,}
\end{equation}

where $g_{k}$ is a pre-determined filter function, and the parameter $N_f$ is the filter half-width. This method consists in filtering the correlation matrix in order to augment the diagonal similarity, thus producing the most coherent modes with specific frequency bands. Such procedure conserves the total energy of the flow redistributing the energy along the frequency spectrum.

The present SPOD methodology is employed due to the interest in finding coherent structures without a previous knowledge of their frequency content. 
Other POD variants have been proposed in the literature. For example, the SPOD approach presented by \citet{towne2018} consists of Fourier-transforming the data in an a priori step. This method was also tested with our datasets but results are not presented for brevity. However, we mention that the spatial support of the coherent structures from this previous method was similar to those obtained with the present one \citep{sieber2016} 
for frequencies in the same spectral band. 

After the filtering procedure of the SPOD, a singular value decomposition (SVD) is performed to the filtered correlation matrix to extract the singular values $\lambda$ and singular vectors $\textbf{A}$, where the modes are ordered by their energy content. The left singular matrix $\textbf{A}$ gives the temporal dynamics of the modes $a_i$ described in equation \ref{eqn:q} as
\begin{equation}\label{eqn:svd}
  \textbf{S}\textbf{A} =  \lambda \textbf{A} \mbox{ .}
\end{equation}
Finally, the spatial modes are computed from the projection of the fluctuation field snapshots into the temporal coefficients as
\begin{equation}\label{eqn:spatial_modes}
  \Phi =  \textbf{Q}\textbf{A} \mbox{ .}
\end{equation}

The SPOD methodology described is applied for the tangential, wall-normal and spanwise velocities for both simulations. The spatial domain used in the computations is restricted to the O-grid along $0.3 \leq x \leq 0.9$ to better capture the dynamics in the developed TBL and avoid the influence of the tripping region and the wake. In the present results, a Gaussian filter is employed since it allows an enhanced control in the SPOD response as shown by \citet{ribeiro2017}. Several filter half-widths were tested and results are presented for a half-width of 30\%. As shown by \citet{ricciardiAIAA}, the snapshot POD provides modes composed of a broad range of frequencies for turbulent boundary layers, and this leads to a poor mode pairing for periodic coherent structures. In these cases, filtering the correlation matrix generates SPOD modes at specific frequency bands, improving the mode pairing \citep{ribeiro2017}.

\begin{figure}
  \centering
    \begin{subfigure}{0.49\textwidth}
    \includegraphics[width=\textwidth]{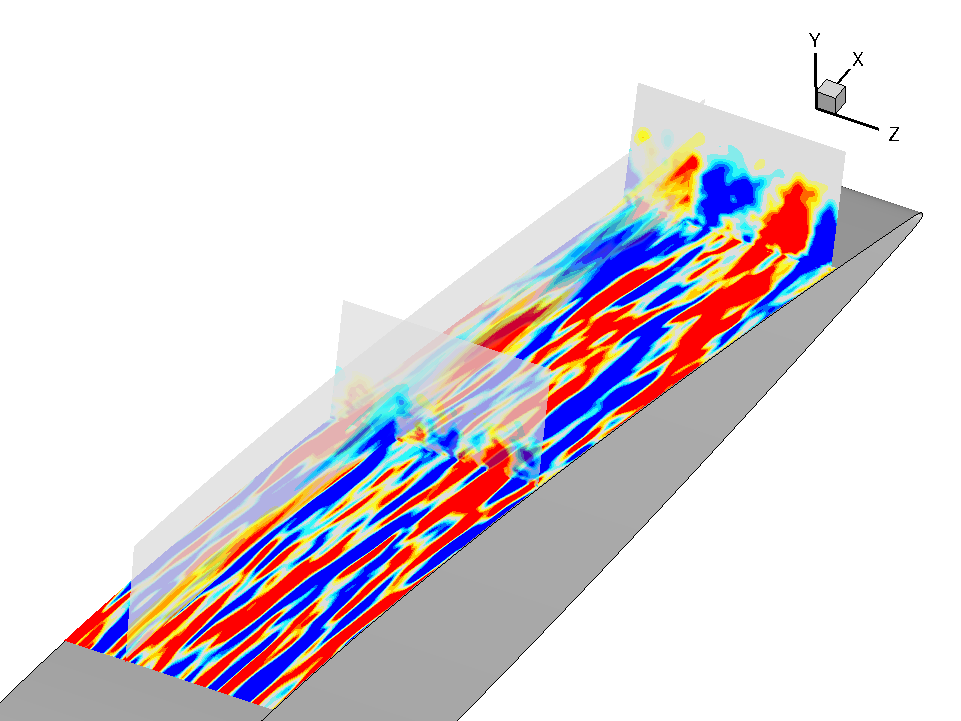}
    \end{subfigure}
    \begin{subfigure}{0.49\textwidth}
    \includegraphics[width=\textwidth]{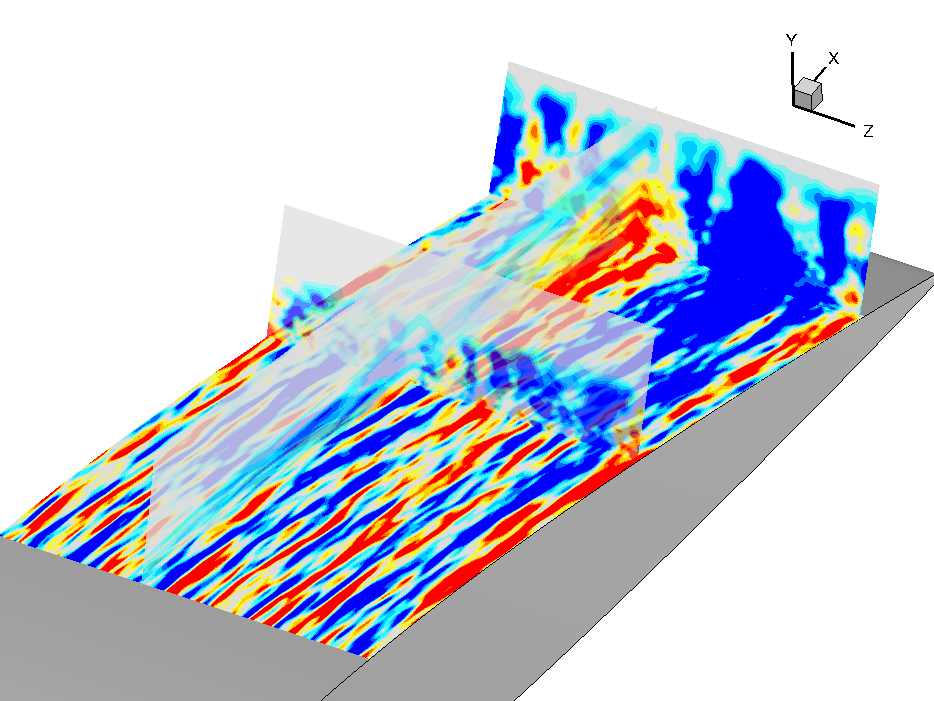}
    \end{subfigure}
  \caption{Longitudinal, near-wall, and transverse planes showing the spatial support of SPOD mode 1 computed in terms of the tangential velocity component ($u_t$) for (\textit{a}) $9$ and (\textit{b}) $12$ deg. angle of attack.}
\label{fig:modes_ut}
\end{figure}

Figures \ref{fig:modes_ut}(\textit{a}) and (\textit{b}) present a near-wall plane depticting the spatial eigenfunctions corresponding to the most energetic SPOD mode for the 9 and 12 deg. simulations, respectively. These results are computed for the tangential velocity $u_t$, where red and blue contours depict positive and negative fluctuations. Longitudinal and transverse planes are also shown to visualize the spatial development of the coherent structures. In both simulations, streamwise elongated structures are observed along the airfoil chord characterizing the presence of near-wall streaks. As the APG strength increases along the airfoil chord, the initially thin and elongated structures become thicker, growing along the entire boundary layer including its outer region. This is evident for the 12 deg. angle of attack case, where a massive low-speed structure with spatial support along the entire boundary layer thickness reaches the trailing edge region. The transverse planes show that the structure has a pronounced spatial growth in the downstream flow region, where the pressure gradient depicts a sharp increase as shown in figure \ref{fig:int_quant}(\textit{b}).

\begin{figure}
  \centering
    \begin{subfigure}{0.49\textwidth}
    \includegraphics[width=\textwidth]{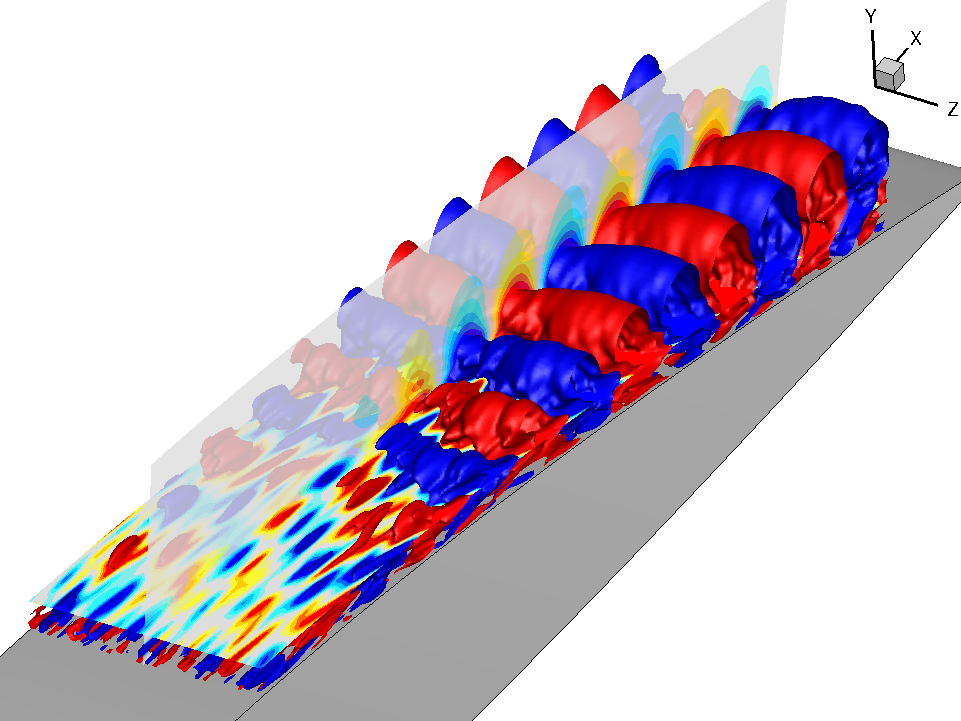}
    \end{subfigure}
    \begin{subfigure}{0.49\textwidth}
    \includegraphics[width=\textwidth]{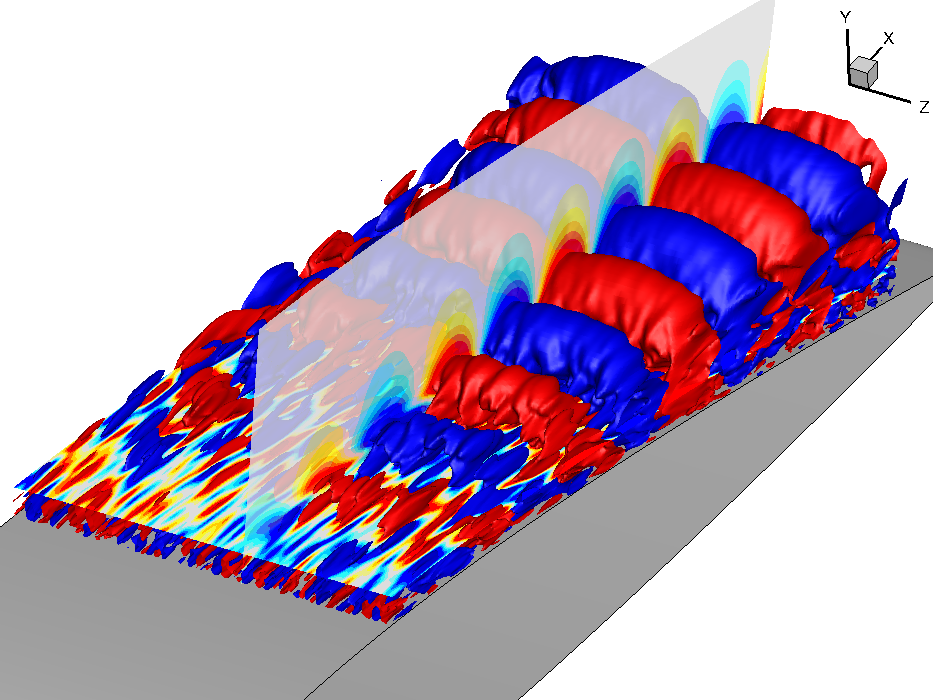}
    \end{subfigure}
  \caption{Isosurfaces, longitudinal and away-from-the-wall planes showing the spatial support of SPOD mode 1 computed in terms of the wall-normal velocity component ($u_n$) for (\textit{a}) $9$ and (\textit{b}) $12$ deg. angle of attack.}
\label{fig:modes_un}
\end{figure}

The SPOD results of the wall-normal velocity component $u_n$ are presented for 9 and 12 deg. in figures \ref{fig:modes_un}(\textit{a}) and (\textit{b}), respectively. Again, the most energetic mode is shown for both cases. These plots show the presence of two-dimensional rollers along the outer layer, resembling the typical structures of a Kelvin-Helmholtz instability.
The coherent structures arise further upstream for the 12 deg. setup when compared to the 9 deg. due to the stronger APGs. The wavelength of these structures is also larger for the higher angle of incidence and, hence, a smaller number of structures is observed. A phase diagram (not shown) computed for the first two SPOD modes shows that the two-dimensional structures are periodic.
The presence of these structures in a region  away from the wall corroborates the arguments of \citet{schatzman2017} that TBLs developing under the influence of APGs may have embedded shear layers. Moreover, for the present flows, the presence of the two-dimensional structures confirm the results from the Rayleigh - Fj{\o}rtoft criteria conducted in section \ref{sec:mean_velocity}, which indicate the presence of an inviscid instability in the outer region of the TBL.

\begin{figure}
  \centering
    \begin{subfigure}{0.49\textwidth}
    \includegraphics[width=\textwidth]{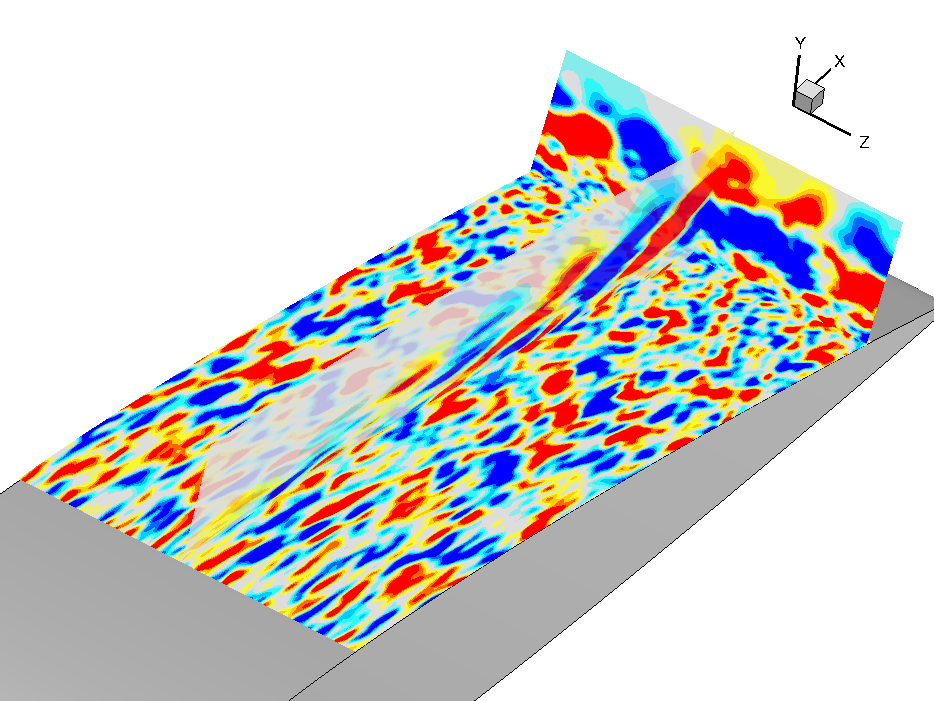}
    \end{subfigure}
    \begin{subfigure}{0.49\textwidth}
    \includegraphics[width=\textwidth]{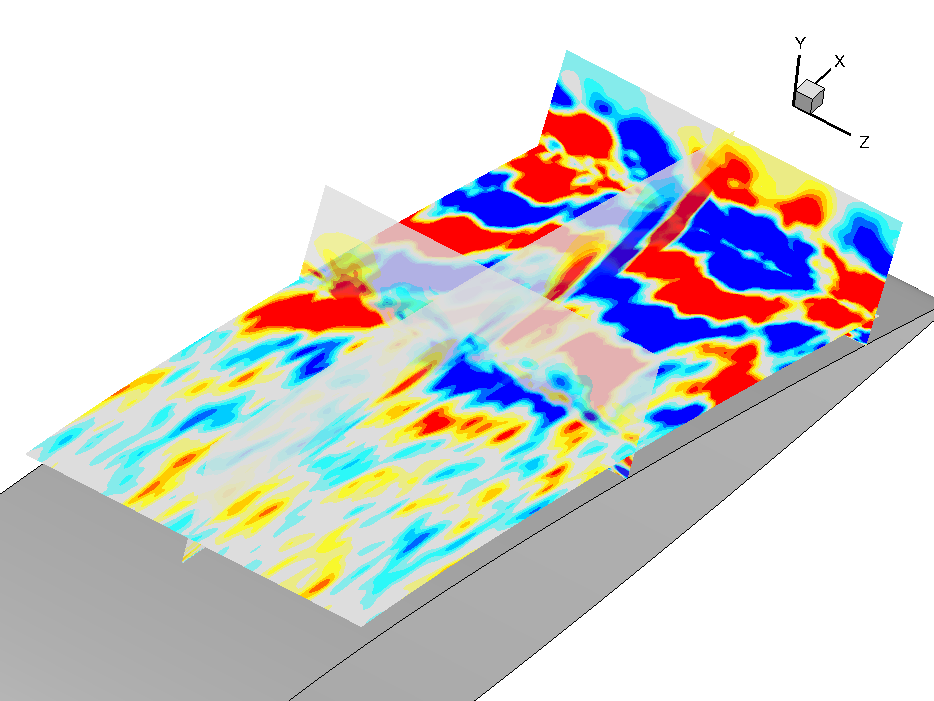}
    \end{subfigure}
  \caption{Longitudinal and transverse planes showing the spatial support of SPOD mode 1 computed in terms of the spanwise velocity component ($w$) for $12$ deg. angle of attack including  (\textit{a}) a near-wall plane, and (\textit{b}) an away-from-the-wall plane.}
\label{fig:modes_w}
\end{figure}

As shown in figure \ref{fig:reynolds}, the $\left< ww \right>$ Reynolds stress components are energized in the outer region of the boundary layer due the APGs. In order to further investigate this behavior, the SPOD analysis is applied to the spanwise velocity component $w$. For brevity, only the most energetic mode for the 12 deg. angle of attack case is presented in figure \ref{fig:modes_w}. Small-scale velocity fluctuations with a more isotropic behavior are observed in the near-wall plane shown in figure \ref{fig:modes_w}\textit{(a)}. However, longitudinal and transverse planes demonstrate that these near-wall structures are ejected from the inner region of the boundary layer along the airfoil, increasing in size at the outer layer. This becomes evident in figure \ref{fig:modes_w}\textit{(b)}, where a plane positioned away from the wall shows the presence of large-scale turbulent structures in the outer region of the boundary layer.
%

\section{Conclusions}

In the present work, wall-resolved LES are applied to investigate the effects of adverse pressure gradients (APGs) on turbulent boundary layers (TBLs) developing on the suction side of a NACA0012 airfoil. The flow Reynolds number is set as $Re = 4 \times 10^5$ and the Mach number is $M = 0.2$. Different APG conditions are studied by changing the airfoil angle of attack, and the present simulations are conducted for incidences of 9 and 12 deg. 
%
The mean velocity profiles show that the APG decelerates the flow near the wall, leading to the occurrence of inflection points in the profiles. 
For moderate and strong APGs, two inflection points appear away from the wall, the first being stable according to Fj{\o}rtoft criterion and the second being unstable. The unstable point indicates the presence of an inviscid instability that originates a shear layer within the TBL. 

The analysis of the Reynolds stress tensor shows that its components are significantly impacted due to the increasing APG for both angles of attack. The tangential stress component presents a secondary peak in the outer region for moderate and strong APGs, 
with the second peak being more pronounced than the inner one at the trailing edge for the 12 deg. angle of attack case. The outer region of all Reynolds stress components are energized by the presence of the APG, which indicates an increase in the turbulence intensity in this region.
A TKE budget evaluation demonstrates that the production term is also considerably affected by the APG. In this context, the inner-layer peak of production is increased with the APG and a plateau is encountered in the outer layer for mild and moderate APGs, while a secondary peak appears for strong APGs. A detailed analysis of the production term demonstrates that its main contribution is due to mean shear, as in ZPG-TBLs. 
However, in the outer region, the production is also affected by other components which depend on the streamwise flow deceleration and the variation of the wall-normal velocity along the boundary layer. 
The dissipation term is also impacted by the APG, increasing throughout the entire boundary layer, especially in the region close to the wall, where it is balanced by the viscous and pressure diffusion terms. 

An analysis of the Lumley triangle and the normalized anisotropy tensor shows that, for mild APGs, the trajectories along the Lumley triangle resemble those of a ZPG-TBL. However, a departure from the standard TBL trajectory is observed as the APG increases. For mild and moderate APGs, a quasi 1D  turbulence state is reached due to the presence of near-wall streaks. On the other hand, for strong APGs, the trajectories along the Lumley triangle never approach the 1D state. For these cases, a quasi 2D isotropic state is reached near the wall since longitudinal and spanwise Reynolds stresses have similar magnitudes. 
For moderate and strong APGs, the secondary peak of the Reynolds stress depicts an axisymmetric contraction state in the Lumley triangle, resembling the trajectory observed for turbulent shear layers.

Flow modal decomposition is performed by SPOD applied to the tangential, wall-normal and spanwise velocity fluctuations. The most energetic SPOD mode from the former shows streaks along the airfoil suction side which grow along the spanwise and wall-normal directions as the APG increases. As also revealed by the SPOD analysis, large-scale, energetic spanwise velocity fluctuations form in the outer region of the boundary layer due to ejection of small-scale structures near the wall. The most energetic SPOD modes of the wall-normal velocity component show two-dimensional structures forming along the embedded shear layer in the outer region of the TBL, corroborating the present stability and flow anisotropy analyses.


\section*{Acknowledgments}

The authors would like to acknowledge Fun\-da\-\c{c}\~{a}o de Amparo \`{a} Pesquisa do Estado de S\~{a}o Paulo, FAPESP, for supporting the present work under research grants No.\ 2013/08293-7, 2021/06448-0 and 2022/00256-4, and Conselho Nacional de Desenvolvimento Científico e Tecnológico, CNPq, for supporting this research under grant No.\ 308017/2021-8.
We also thank SDUMONT-LNCC (Project SimTurb) and CENAPAD-SP (Project 551) for providing the computational resources used in this work.

 \bibliographystyle{elsarticle-harv}
 \bibliography{refs}



\end{document}